\documentclass[largeformat]{interact}

\usepackage[caption=false]{subfig}

\usepackage{amsmath}
\usepackage{mathtools}
\usepackage[normalem]{ulem}
\usepackage[ruled,vlined,linesnumbered]{algorithm2e}
\usepackage{tikz}
\usepackage{tikzscale}
\usepackage[numbers,sort&compress,merge]{natbib}
\bibpunct[, ]{[}{]}{,}{n}{,}{,}

\usepackage{color}
\definecolor{etienne}{rgb}{0.7, 0.11, 0.11}

\newcommand{\pd}{d}

\newcommand{\R}{\mathbb{R}}

\newcommand{\cEp}{\mathcal{E}_{\sf p}}

\renewcommand{\sp}{{\sf p}}
\newcommand{\sq}{{\sf q}}
\newcommand{\piv}{{\sf piv}}
\newcommand{\tsC}{\widetilde{{\sf C}}}
\newcommand{\sB}{{\sf B}}
\newcommand{\sC}{{\sf C}}

\newcommand{\sCp}{{\sf C}_{\sp}}
\newcommand{\tsCp}{\widetilde{{\sf C}}_{\sp}}

\newcommand{\sD}{{\sf D}}
\newcommand{\tsD}{\widetilde{\sf D}}
\newcommand{\sDp}{{\sf D}_{\sp}}
\newcommand{\tsDp}{\widetilde{\sf D}_{\sp}}

\newcommand{\sEp}{{\sf E}_{\sp}}

\newcommand{\sFp}{{\sf F}_{\sp}}

\newcommand{\sGp}{{\sf G}_{\sp}}

\newcommand{\shp}{{\sf h}_{\sp}}
\newcommand{\sL}{{\sf L}}

\newcommand{\sSp}{{\sf S}_{\sp}}

\renewcommand{\sGp}{{\sf G}_{\sp}}
\newcommand{\sU}{{\sf U}}
\newcommand{\sV}{{\sf V}}

\DeclareMathOperator{\Tr}{Tr}

\DeclareMathOperator{\Span}{Span}

\DeclareMathOperator*{\argmin}{arg\,min}
\DeclareMathOperator{\tr}{T}

\newcommand{\GrLog}{{\rm Log}_{\rm Gr,0}}
\newcommand{\GrExp}{{\rm Exp}_{\rm Gr,0}}

\newcommand{\Nb}{N_{\sf b}}

\newcommand{\Ma}{\mathcal M}

\newcommand{\Gr}{\mathcal M_{\rm Gr}}

\DeclarePairedDelimiterX\Set[2]{\lbrace}{\rbrace}{ #1 \,\delimsize|\, #2 }

\newcommand{\black}{\color{black}}

\newcommand{\gd}[1]{{\small \color{blue} #1}}

\newcommand{\mathsout}[1]% will draw line through middle of #1
{\bgroup\mathchoice
  {\sbox0{$\displaystyle{#1}$}%
    \usebox0\hspace{-\wd0}%
    \rule[0.5\ht0-0.5\dp0-.5pt]{\wd0}{1pt}}%
  {\sbox0{$\textstyle{#1}$}%
    \usebox0\hspace{-\wd0}%
    \rule[0.5\ht0-0.5\dp0-.5pt]{\wd0}{1pt}}%
  {\sbox0{$\scriptstyle{#1}$}%
    \usebox0\hspace{-\wd0}%
    \rule[0.5\ht0-0.5\dp0-.5pt]{\wd0}{1pt}}%
  {\sbox0{$\scriptscriptstyle{#1}$}%
    \usebox0\hspace{-\wd0}%
    \rule[0.5\ht0-0.5\dp0-.5pt]{\wd0}{1pt}}%
\egroup}

\begin{document}

\title{An approximation strategy to compute accurate initial density matrices for repeated self-consistent field calculations at different geometries}

\author{
  \name{
    \'E.~Polack\textsuperscript{a}
    A.~Mikhalev\textsuperscript{b}
    G.~Dusson\textsuperscript{a}   
    B.~Stamm\textsuperscript{b}
    and F.~Lipparini\textsuperscript{c}
  }
  \affil{
    \textsuperscript{a}Laboratoire de Math\'ematiques de Besan\c{c}on, UMR CNRS 6623, Universit\'e Bourgogne Franche-Comt\'e, 
    16 route de Gray, 25030 Besan\c{c}on, France\\
    \textsuperscript{b}Center for Computational Engineering Science, RWTH Aachen University, Schinkelstr. 2, 52062 Aachen, Germany\\
    \textsuperscript{c}Dipartimento di Chimica e Chimica Industriale, Univerist\`a di Pisa, Via G. Moruzzi 13, I-56124 Pisa, Italy
  }
}

\maketitle

\begin{abstract}
Repeated computations on the same molecular system, but with different geometries, are often performed in quantum chemistry, for instance, in ab-initio molecular dynamics simulations or geometry optimizations. While many efficient strategies exist to provide a good guess for the self-consistent field procedure, which is usually the main computational task to be performed, little is known on how to efficiently exploit in this direction the abundance of information generated during the many computations. In this article, we present a strategy to provide an accurate initial guess for the density matrix, expanded in a set of localized basis functions, within the self-consistent field iterations for parametrized Hartree-Fock problems where the nuclear coordinates are changed along a few user-specified collective variables, such as the molecule's normal modes. 
Our approach is based on an offline-stage where the Hartree-Fock eigenvalue problem is solved for some particular parameter values and an online-stage where the initial guess is computed very efficiently for {\it any} new parameter value.
The method allows non-linear approximations of density matrices, which belong to a non-linear manifold that is isomorphic to the Grassmann manifold.
The so-called Grassmann exponential and logarithm map the manifold onto the tangent space and thus provides the correct geometrical setting accounting for the manifold structure when working with subspaces rather than functions itself.
Numerical tests on different amino acids show promising initial results.
\end{abstract}

\begin{keywords}
Self-consistent field, density guess, ab-initio molecular dynamics, geometry optimization
\end{keywords}

\section{Introduction}
Computational quantum chemistry allows nowadays to describe, model and predict a very large variety of chemical phenomena. Thanks to a combination of new methods, computational techniques and hardware developments, quantum chemistry can be used to compute molecular structures, spectroscopic and response properties, reaction paths, aggregation properties and much more.
A typical computational setup starts usually with the prediction, at a given level of theory, of the molecular geometry, which is obtained by minimizing the Born-Oppenheimer energy with respect to the nuclear coordinates~\cite{Schlegel_WIREs_GeoOpt}. Properties calculations are then carried out. For large molecules, as several stable conformers can exist, these operations may need to be repeated in order to account for the existence of multiple minima. The number of calculations required can be further increased if more complex systems are considered, for instance, a large biological polymer or a solvated molecule, as a correct statistical sampling of the system's configurations becomes mandatory in order to achieve correct results. In such cases, calculations can be performed on snapshots taken from classical or ab-initio molecular dynamics. As a consequence, a computational study often requires to perform several calculations on the same system at different geometries.

One of the most common task performed during a quantum chemical calculation is the solution to the self-consistent field (SCF) equations, that is at the basis of Hartree-Fock~\cite{Roothaan_RMP_SCF} (HF) and Density Functional Theory~\cite{Kohn_PR_DFT} (DFT). The latter can often be the method of choice for the overall computational study, while the former is at least a necessary starting point for more refined post-HF treatments. The SCF equations are a set of coupled, non-linear differential equations that are solved iteratively. As such equations can exhibit notorious convergence problems~\cite{Schlegel_SCFConv}, in the last years a number of different numerical techniques have been developed to achieve reliable and fast convergence. These new developments include not only convergence acceleration techniques, such as the popular Direct Inversion in the Iterative Subspace~\cite{Pulay_CPL_DIIS,Pulay_JCC_DIIScom} (DIIS) and its many extensions and generalizations~\cite{Cances_JCP_EDIIS,Wang_JCP_ADIIS}, but also methods to provide a better guess to the iterative procedure~\cite{Wolfsberg_JCP_Guess,Hoffmann_JCP_ExHuc,Almlof_JCC_SAD,Harris_PRB_Guess,VanLenthe_JCC_SAD,Lethola_JCTC_Guess,Lehtola_JCP_SAP}.  
The latter point is of particular importance, as the SCF procedure can be particularly problematic when starting from an unrealistic guess and exhibit large oscillations and other pathological behaviors~\cite{Lethola_JCTC_Guess}. Thanks to all these recent developments, many existing SCF implementations manage to achieve convergence,
at least for closed-shell systems, in as little as 15-20 iterations. 

The guessing procedure developed in the years for SCF are usually focused on providing a good estimate of the electronic density for single point calculations. Much less has been done to specifically address the issue of repeated calculations, 
other than common-sense practices, such as using the density of a previous point as a guess for the next energy and forces evaluation in a geometry optimization and other related strategies. 
A notable and particularly successful exception, that directly aims at providing a better guess for the SCF procedure in ab-initio MD simulations (AIMD), is based on extended-Lagrangian techniques~\cite{Niklasson_XLBOMD,Loco_JCTC_QMMD}, which introduce an auxiliary density that is propagated along the dynamics and used to provide a guess that is usually sufficiently good, so that, at the precision required by AIMD simulations, only a few SCF iterations are required per step. 
These techniques, that use the density, or guess density, at a collection of previous steps (usually, from a couple to about ten steps), successfully exploit this information to improve the guessing procedure. 
However, extended-Lagrangian techniques rely on the fact that the nuclei configurations at the various steps are produced by a deterministic process, such as MD, and are therefore not applicable to a general repeated calculation, as in geometry optimizations or QM/MM snapshots originating from uncorrelated frames extracted from a MD simulation.

In this work, we try to address the problem of forming a guess for repeated calculations that is as general and robust as possible. In particular, our aim is to develop a procedure that is able to reuse as much information as possible from previous calculations at different geometries, independent of their provenance, to provide an optimal guess for a further calculation. We assume that a set of atom-centered, localized basis functions, such as gaussian-type orbitals, is employed. 
The main idea can be stated as follows. Let us consider a set of configurations for which the SCF density is known and a further, new configuration for which we want to guess the density. A naive strategy would be to linearly interpolate the configurations, i.e., their Cartesian coordinates, for instance, and apply the same interpolation to the density matrices. However, there would be no guarantee that the density obtained with such a procedure would indeed be a density matrix, stemming from a monodeterminantal wavefunction. In order to enforce the correct properties of the new, approximated density, we adopt a geometric perspective. From a mathematical point of view, the density matrices live in a so-called Grassmann Manifold which, as it is not a vector space, does not allow for linear interpolation to be used. However, we will show how it is possible to map a point in such a manifold to its tangent space, which indeed is a vector space, perform the interpolation, or any kind of approximation there, and then go back to the manifold, ensuring that the interpolated density has all the properties that are required for it to be a genuine density matrix. 

The techniques that we use are conceptually related to 
notions that are not new to chemists. 
Indeed, it is known that orbital rotations can be parametrized in terms of exponential maps, and that such maps can be used to parametrize the effect of orbital rotations on the density matrix. 
This is commonly done for direct orbital optimization techniques, used for quadratically convergent SCF~\cite{Douady79,Bacskay_CP_QCSCF} and multiconfigurational SCF implementations~\cite{Werner81,Jensen84a,Jensen84b,Werner85,Jensen86,Lipparini16}. In this contribution, we use a different notion of exponential which allows to efficiently parametrize the set of density matrices. However, these exponentials are in practice very different, due to the structure difference between orbital rotations and density matrices.
By applying geometrical techniques to the problem of repeated calculations, we will show how a very effective and rigorous SCF guessing procedure can be developed.

On the other hand, solving problems repeatedly for different parameter values is common in many engineering applications and can be put under the context of many-query computations. In such scenario, the concept of reduced order modelling for parametrized problems has been established and it has become a mature tool in computational engineering science. 
The roots of modern reduced order modelling lie in structural mechanics and an overview of literature, methods, concepts and applications can be found in the monograph~\cite{hesthaven2016certified}.
The concept of reduced order modelling is only little known and exploited in computational chemistry. The few contributions in this field \cite{cances2002towards,cances2007feasibility,maday2008reduced,schauer2015reduced} involve methods based on finite elements, with only a limited amount of work having been done for Gaussian-type atomic orbitals.
It can be noted that the numerical results in these papers deal with rather small molecules and do not contain any geometrical considerations as presented in this work.
We hope that our further contribution shades a different angle at reduced order modelling for parametrized problems in electronic structure calculation.

In this preliminary study, we develop the methodology and apply the newly developed technique to a simple problem, where we assume that no level crossing occurs between the states due to geometry displacements. In particular, we generate one- and two-dimensional grids of molecular geometries by displacing the equilibrium geometry of a few chosen molecular systems along one resp. two different normal coordinates, using displacements of up to one atomic unit times the normalized coordinates. While this is a very simplified problem with respect to the general one, it provides an example of small, but non negligible oscillations of the geometry around an equilibrium point that are typical of MD simulations or of anharmonic force field calculations. We show that using a small number of data, we are able to predict the density at all other points with remarkable accuracy, providing an almost already converged density matrix.

This paper is organized as follows.
In Section~\ref{sec:pb_statement}, we describe the addressed problem, namely the development of good initial guesses for the solution of the SCF problem parametrized with respect to the atomic positions, and we present the corresponding equations.
We then present the methodology in Section~\ref{sec:met}, starting in Section~\ref{sec:geom} with the geometrical structure of the object of interest: the density matrix. We continue by describing the process of computing an approximation of the density matrix in Section~\ref{sec:approx_DM}, first in a case where the parameter dependency is one-dimensional, and second in the more complicated case of a multi-dimensional parameter space.
In Section~\ref{sec:numerics}, we present some numerical results illustrating the accuracy of the initial guesses as well as the low computational cost obtained by this method.
We close this article by pointing out some perspectives in Section~\ref{sec:perspectives}.

\section{Problem statement}
\label{sec:pb_statement}
While there exists a map between the geometry of a molecule and, for a given basis set, its SCF density matrix, such a map is unknown and certainly highly nonlinear.
Finding the exact approximation of this map seems thus an impossible task.
We have therefore to resort to some kind of approximation.
The problem that we want to address in this article can be stated as follows. Suppose that a set of SCF computations has to be performed on the same molecule, or cluster of molecules, at different geometries, for instance, in a geometry optimization or molecular dynamics simulation. 
Suppose also that we allow the problem to be solved at some few specific geometries in order to access the density matrices for those points.
The following question arises: how can the pre-computed density matrices be used to approximate the solution at any new point, or to provide a very robust guess for the SCF at this new point?

To address this problem, a strategy needs to be developed in order to actually define geometries where we first compute the density matrix and then, in a second phase, use them to provide a guess and thus this task has the flavour of an interpolation or more generally an approximation problem.
However, this is not an easy task due to the fact that the SCF density matrices are not elements of a vector space. This means that in general, a linear combination of two density matrices is not a density matrix. 
Therefore, the first goal of this paper is to find a strategy to perform an interpolation, or more generally an approximation, of the available densities in the appropriate set (manifold), so that the resulting density has all the properties that are needed.

A further point concerns the overall efficiency of this process, that strongly depends on how much data is needed to get a good approximation to the density. 
In other words, if we need to solve the SCF equations for a large number, say $N_{\rm g}$, different geometries, we want to be able to provide a good guess based on pre-computed density matrices at $Q \ll N_{\rm g}$ points. 
Therefore, the second problem that we want to address is how can one find a \emph{minimal} number of points that allow one to build a good density approximation at all other points.

Let us start by stating the first problem in a more precise way. We consider the electronic Schr\"odinger problem where the $M$ nuclear positions $\bm r\in \mathbb R^{3M}$ are parametrized by a given, possibly non-linear, map $\mathbb P \ni \sp\mapsto \psi(\sp) =\bm r\in \mathbb R^{3M}$, where the map $\psi$ may consist of reaction-coordinates, optimization steps, normal modes or any collective variable in general. 
We refer to the bounded domain $\mathbb P\subset \R^P$, for a given $P\in\mathbb N$, as the parameter domain. The parameter-dependency plays a key role in the methodology and we therefore highlight the dependency on $\sp$  in the following with a subscript.
We consider a level of theory that corresponds to the Hartree-Fock equations or Density Functional Theory (DFT) but without loss of generality we present in the following our approach for the Hartree-Fock (HF) method. 
Using a given basis set within the LCAO-framework (Linear Combination of Atomic Orbitals), the discrete energy can be written as
\begin{equation}
  \label{eq:hf}
  \cEp(\tsC) = \Tr \left( \tsC^{\top}\shp \tsC + \tfrac12\tsC^{\top}\sGp(\tsC\tsC^{\tr})\tsC \right)
\end{equation}
where $\shp$ and $\sGp$ are the customary one and two electron integral matrices in the atomic orbitals (AO) basis for the parameter value $\sp$. The matrix $\tsC\in\R^{\Nb\times N}$ contains the $\Nb$ coefficients of the $N$ occupied molecular orbitals within the given $\Nb$-dimensional basis.
The SCF problem can be stated as the variational minimization of the SCF energy
\begin{equation}
  \min_{\tsC\in\Ma(\sp)} \cEp(\tsC),
\end{equation}
where the coefficients $\tsC$ need to satisfy the usual orthonormality constraints or, in other words, belong to the manifold $\Ma(\sp)$ defined as
\begin{equation}
	\Ma(\sp) = \left\{ \tsC \in \R^{\Nb \times N} \;\middle|\; \tsC^{\top}\sSp \tsC = 1_{N} \right\},
\end{equation}
with $\sSp$ denoting the overlap matrix.
Writing the first-order optimality conditions, we obtain the following non-linear eigenvalue problem:
Find a matrix \( \tsCp \in \Ma(\sp)\) and a diagonal matrix
\( \sEp \in \R^{N\times N} \) containing the orbital energies~\( (\varepsilon_{1}, \ldots, \varepsilon_{N}) \) such that
\begin{align}
  \label{eq:evp1}
  \sFp(\tsDp)\tsCp &= \sSp \tsCp \sEp, \\
  \label{eq:evp2}
  \tsCp \tsCp^{\top} &= \tsDp, 
\end{align}
where \( \sFp(\tsD) = \shp + \sGp(\tsD) \) denotes the Fock operator and \( \tsDp \in \R^{\Nb\times \Nb} \) the density matrix.
We note that the input data $\sFp$ and $\sSp$ depend explicitly on $\sp$ whereas the solution $\tsCp$ respectively~$\tsDp$ to the eigenvalue problem depend implicitly on $\sp$ through the relations \eqref{eq:evp1}--\eqref{eq:evp2}.

When the computation is done without any previous history (single step calculation as
opposed to molecular dynamics, for example, where a predictor can be employed),
an initial guess contains no a priori information on the solution and provides an error of order one.
As already mentioned, the goal of this paper is to establish an approximation scheme to provide a good guess of the density matrix $\tsDp$ when some known density matrices $\left\{\tsD_{\sp_{i}} \right\}_{i=1}^{Q}$ for some parameter values $\left\{ \sp_{i} \right\}_{i=1}^{Q}$ are given. 
Our strategy tackles the two main issues stated at the beginning of this section as follows.

\begin{figure}[t!]
	\centering
    \includegraphics[trim = 10mm 190mm 30mm 20mm, clip,width=0.4\textwidth]{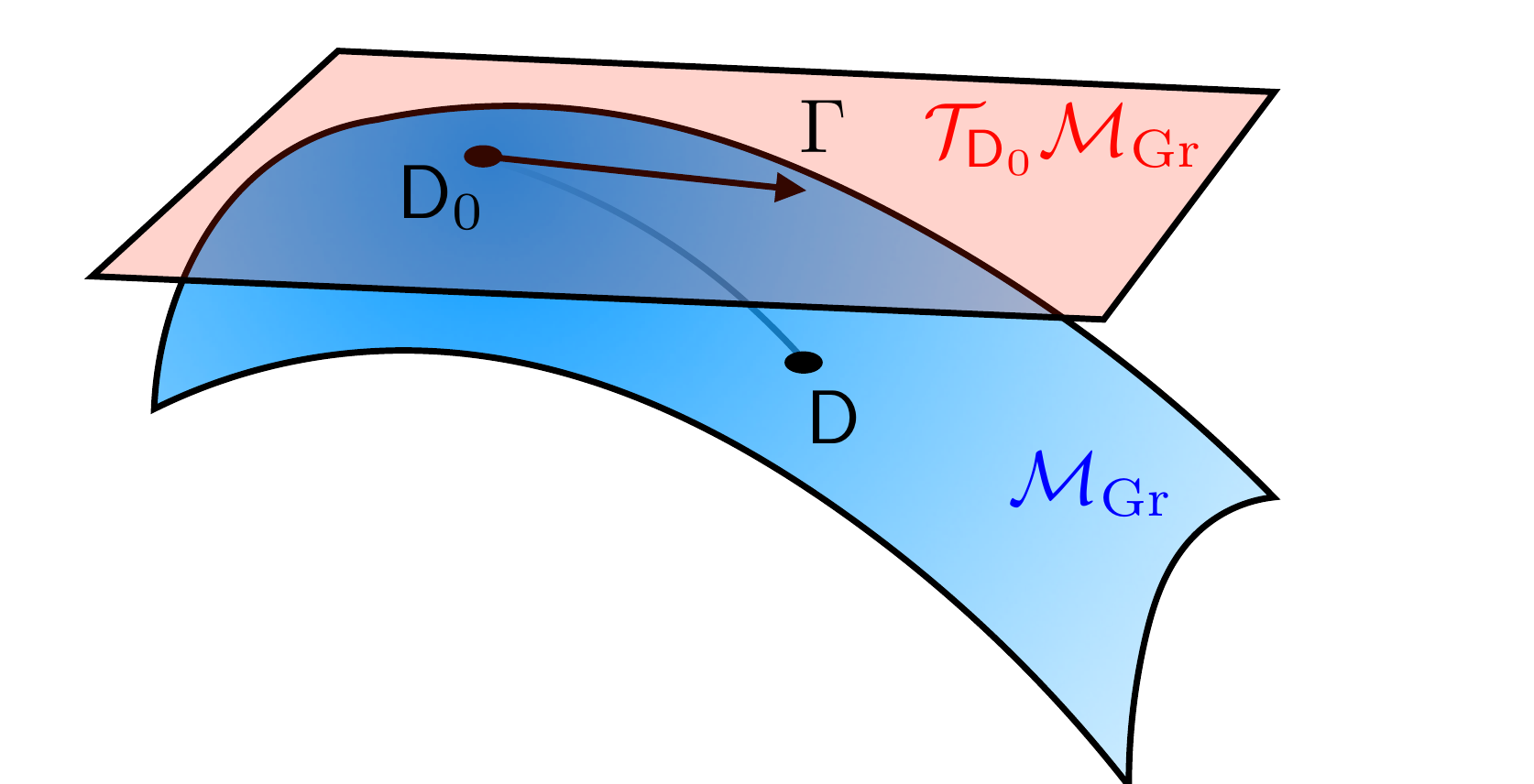}
    \includegraphics[trim = 10mm 190mm 30mm 20mm, clip,width=0.4\textwidth]{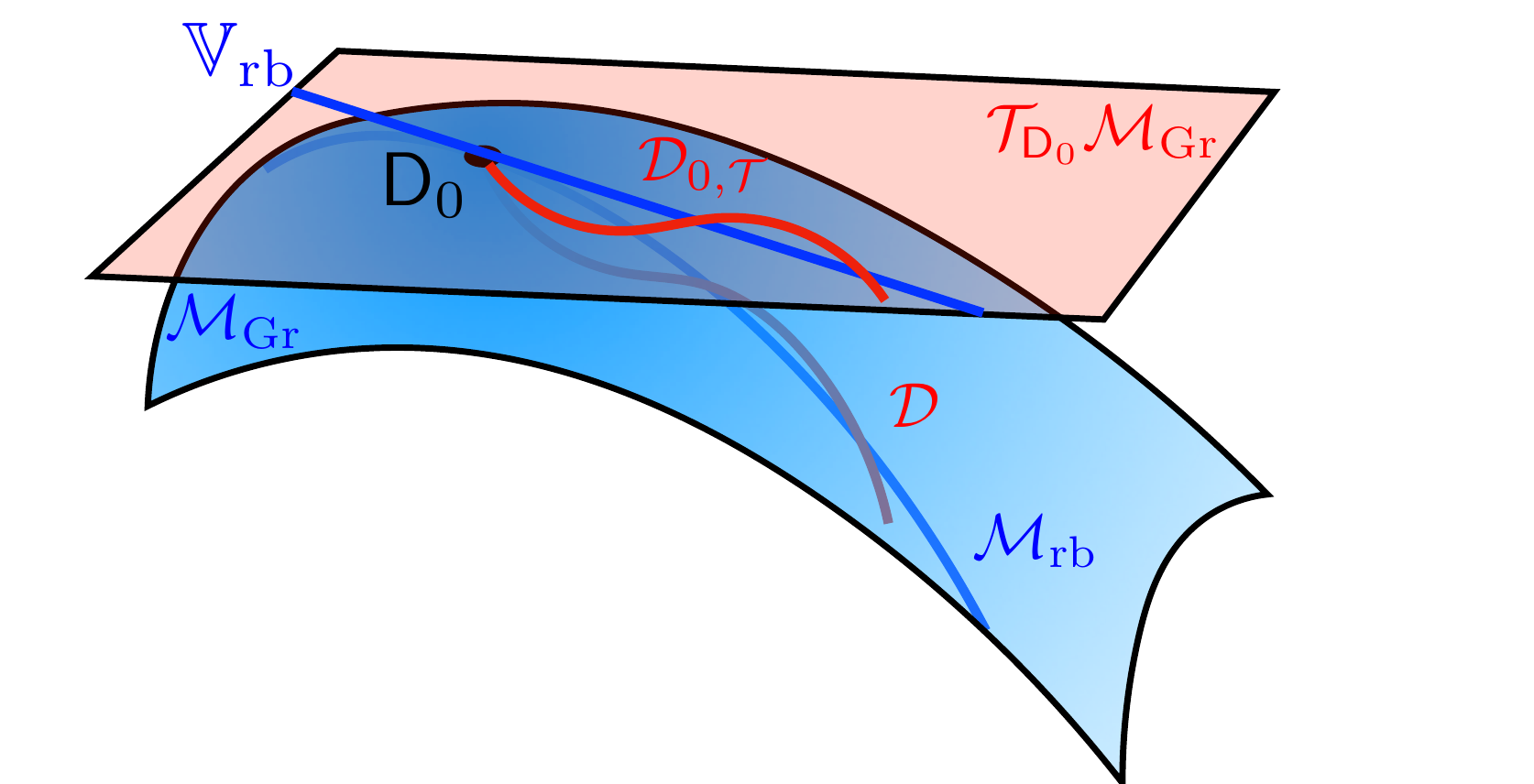}
	\caption{ Schematic illustration of the geometrical setting. In both figures, we illustrate by the blue hypersurface the Grassmann manifold $\Gr$ and by the red plane the tangent space $\mathcal T_{\sD_0}\Gr$ to $\Gr$ at $\sD_0$. On the left, we illustrate the one-to-one relationship between a close density matrix $\sD\in\Gr$ and the corresponding vector  $\Gamma=\GrLog \sD$ in the tangent space.
	On the right, we further schematically illustrate the notions of $\mathcal D_{0,\mathcal T}$ and $\mathbb V_{\rm rb}$, respectively defined in \eqref{eq:DT0} and \eqref{eq:VRB}, as well as their equivalent sets $\mathcal D=\{\sD_\sp|\sp\in\mathbb P\}$ and $\mathcal M_{\rm rb}=\GrExp(\mathbb V_{\rm rb})$ on $\Gr$.
}
	\label{fig:Gr}
\end{figure}

First, in order to be able to perform an approximation based on known densities, we look at the problem from a geometrical point of view. 
The orthogonal projectors onto the space spanned by the $N$ orbitals in the atomic orbital basis belong to the manifold
\begin{equation}
    \label{eq:GrassManif}
	\sSp^\frac12\tsDp\sSp^\frac12 \in \Gr = \left\{ \sD  \in \real^{\Nb \times \Nb}\;\middle|\; \sD=\sD^\top, \sD^2=\sD, \Tr(\sD) = N  \right\},
\end{equation}
is well known in mathematics under the name ``Grassmann manifold".  
To be completely rigorous, the former is isomorphic to the latter, but we omit such technical details in the following. 
 Here, using the properties of such manifold, we develop a strategy that maps the densities obtained at the various points to a vector space, namely the tangent space, performs the linear approximation there and then maps the interpolation back to the Grassmann manifold. From an intuitive point of view, the process can be seen as depicted in Figure \ref{fig:Gr} (left). 
The differentiable manifold can be thought of as a curve hypersurface (in blue) of lower dimension.
We map the manifold to the hyperplane, which is tangent to the surface at a given point,  and are projecting then all the data points (density matrices) to such plane and perform the approximation. 
 Then, once the approximation is built, we use the inverse map and go back to the manifold. The key point here is that this guarantees to obtain a density matrix that satisfies all the physical requirements and we are therefore sure that such a matrix corresponds to a single Slater Determinant.

Second, we address the computational problem of making this scheme efficient. Indeed, if one computes the density matrix for a large number of molecular geometries, it is very likely that the information will be redundant. 
In this case, the corresponding density matrices can all be approximated by (different) linear combinations of very few common elementary matrices. In applied mathematics, those elements are called reduced basis of the parametrized problem since they build a basis of a vector space that approximates any density matrix to high accuracy. 
As elaborated above, the issue with this approach in our context is that any
linear combination of density matrices is in general not a density matrix.
However, we can apply this concept on the tangent space. Thus, after having
mapped all density matrices to the tangent space, one can, for example, find a
low dimensional basis by performing a singular value decomposition (SVD) of all
tangent vectors. In consequence, any tangent vector can be represented with few
degrees of freedom if expressed in this ``reduced basis'' on the tangent space.
Mapping this approximation back to the manifold of density matrices guarantees
then that the approximation has the structure of a density matrix.

\section{Methodology}
\label{sec:met}
\subsection{The geometrical structure}
\label{sec:geom}
We note that for any value of the parameter $\sp$, the matrix $\tsCp\in\Ma(\sp)$, solution to~\eqref{eq:evp1}--\eqref{eq:evp2}, can be transformed and we define $\sCp:= \sSp^\frac12 \tsCp$.
In consequence, we observe that $\sCp$ belongs to the Stiefel manifold of orthonormal $N$-frames in $\R^{\Nb}$.
The corresponding density matrix 
$
	\sDp =  \sCp \sCp^\top,
$
belongs to the manifold of rank $N$ projectors in $\R^{\Nb}$, already defined in eq. \eqref{eq:GrassManif},
which is isomorphic to the Grassmann manifold, hence designated with the same name.
We will not insist on a very precise description of the setting in terms of differential geometry as this is not the purpose of this article. For interested readers we refer to~\cite{edelman1998geometry,zimmermann2019manifold}.
We will rather point out the practically important considerations, give some intuitive explanations and try to keep technical considerations to a minimum.
We note that the energy $\cEp$ defined in~\eqref{eq:hf} is invariant under orthogonal transformation of the $N$-frames and we thus conclude that the solution of~\eqref{eq:evp1}--\eqref{eq:evp2} is uniquely represented by $\sDp$ rather than $\sCp$.

We are thus facing the situation where we are given the possibility to access the density matrix $\sDp$ for specific parameter values $\sp$, but we would like to keep those computations to a minimum.
This will be done in the so-called {\it offline}-stage, where two tasks will be assigned. 
First, the choice of the points $\left\{ \sp_{i} \right\}_{i=1}^{Q}$ and second, the computation of the density matrix $\left\{ \sD_{\sp_{i}} \right\}_{i=1}^{Q}$ at each of those points.

In the {\it online}-stage, we are then given parameter-solution pairs $\left\{ \sp_{i}, \sD_{\sp_{i}} \right\}_{i=1}^{Q}$ with $\sD_{\sp_{i}}\in\Gr$ and we aim to approximate the mapping
\begin{equation}
	\label{eq:map}
	\mathbb P \ni \sp \mapsto \sDp\in\Gr.
\end{equation}
Since the Grassmann manifold is not a vector space, it is obvious that a linear combination of density matrices does not belong to $\Gr$ in general.
In consequence, approximating $\Gr$ with a vector space does not respect the geometric structure of the problem and some of the properties of $\Gr$ would be lost in general.

For the Grassmann manifold, which is a differential manifold, for any given $\sD_0=\sC_{0} \sC_{0}^\top$  with $\sD_0:=\sD_{\sp_0}$ and $\sC_0:= \sC_{\sp_0} $ for fixed $\sp_0$, the tangent space is
\begin{equation}
	\mathcal T_{\sD_0}\Gr
	= \left\{ \Gamma \in \R^{\Nb\times N}  \,\middle|\, \sC_{0}^\top \Gamma=0   \right\} 
	\subset \R^{\Nb\times N}.
\end{equation}
Note that the tangent space is an affine space.
One can then introduce the Grassmann exponential which maps tangent vectors on $\mathcal T_{\sD_0}\Gr$ to the manifold $\Gr$ in a locally bijective manner around $\sD_0$.
Indeed, it is not only an abstract tool from differential geometry, but it can be computed in practice involving the matrix exponential.
By complementing $\sC_0$ with orthonormal columns to obtain $(\sC_0,\sC_{\!\perp})\in O(\Nb)$ and $\Gamma\in \mathcal T_{\sD_0}\Gr$ we have
\begin{equation}
    	\GrExp(\Gamma) = \sC \sC^\top,\qquad
        \sC = \begin{pmatrix} \sC_0, \sC_{\!\perp}  \end{pmatrix} \, \exp
        \begin{pmatrix}
            0 & -\sB^\top \\ \sB & 0  
        \end{pmatrix}
        \mathsf I_{\Nb,N},
\end{equation}
where the matrix $\sB\in \mathbb{R}^{(\Nb-N)\times N}$ contains expansion coefficients of  columns of $\Gamma$ in a span of columns of $\sC_{\!\perp}$ such that $\Gamma = \sC_{\!\perp} \sB$ and $\mathsf I_{\Nb,N}=(\mathsf I_N,0)^\top \in \R^{\Nb\times N}$ are the first $N$ columns of the $\Nb\times \Nb$ identity matrix. As one can see, it is an exponential ansatz of a skew-symmetric matrix that leaves out any redundant parametrization (the zero diagonal blocks) due to the mixing of the virtual resp. occupied orbitals. 
In this manner the mapping between $\mathcal T_{\sD_0}\Gr$ and $\Gr$ becomes locally bijective. 
Further, the Grassmann exponential can then be expressed by 
\begin{equation}
	\GrExp(\Gamma) = \sC \sC^\top,\qquad
	\sC
	=
 	\left[ \sC_0 \sV_e \cos(\Sigma_e) + \sU_e \sin(\Sigma_e) \right] \sV_e^\top,
\end{equation}
by means of a singular value decomposition (SVD) $\Gamma = \sU_e \Sigma_e  \sV_e^\top $ of $\Gamma$.
A schematic representation can be found in Figure~\ref{fig:Gr} (left) and we refer  to \cite{edelman1998geometry,zimmermann2019manifold} for further details and its derivation.

The inverse function is the so-called Grassmann logarithm $\GrLog$ (see, e.g.,~\cite{edelman1998geometry,zimmermann2019manifold}) which maps any $\sD = \sC\sC^\top\in\Gr$ in a neighborhood of $\sD_0$ to the tangent space $\mathcal T_{\sD_0}\Gr$ by
\begin{equation}
	\GrLog(\sD) 
	= 
	\sU_{\ell} \arctan(\Sigma_{\ell}) \sV_{\ell}^\top,
\end{equation}
using the following SVD decomposition
\begin{equation}
	\sU_{\ell}\Sigma_{\ell} \sV_{\ell}^\top = \sL
	\qquad 
	\mbox{with} 
	\qquad 
	\sL 
	=\sC \left(\sC_{0}^\top \sC\right)^{-1} - \sC_{0}.
\end{equation}

Note that we respectively denote by $\sU_\ell \Sigma_\ell \sV_\ell^\top$ and $\sU_e \Sigma_e \sV_e^\top$ the thin Singular Value Decompositions (SVD) of $\sL$ and $\Gamma$ with the asymptotic cost of $\mathcal O(\Nb N^2)$, see e.g.~\cite{edelman1998geometry,zimmermann2019manifold}.
Such a cost is comparable with the cost of a traditional dense diagonalization, which is commonly used in SCF codes working with localized basis functions. We remark here that the diagonalization itself is seldom the rate-determining step for medium-large calculations, which are dominated by the cost of building the Fock matrix.
\black

In this manner we map each density matrix $\sD_{\sp_{i}}$ to the tangent space
at the reference point~$\sD_0$ in order to obtain $\Gamma_i = \GrLog(\sD_{\sp_i})$. 
The reference point can in principle be chosen arbitrarily but it is the most intuitive to place it in the center of the parameter domain $\mathbb P$.
Since the tangent space is a vector space we have now transformed our problem to a standard approximation problem of pairs of data $(\sp_i,\Gamma_i)$ belonging to Euclidian vector spaces.
In the next sections, we will precise how the map $\mathbb P \ni \sp\mapsto \Gamma(\sp) \in \mathcal T_{\sD_0}\Gr  $ is approximated.

Before that, we summarize the global picture of our strategy: using the Grassmann logarithm allows us to map density matrices on the tangent space at a particular point of the manifold. 
Then we can rely on classical approximations techniques between the parameter domain and the tangent space being a vector space.
Having the approximation defined on the tangent space, we use the Grassmann exponential to map back to the Grassmann manifold and thus can provide a density matrix obeying the exact geometrical structure of the problem, i.e. belonging to $\Gr$.

\subsection{Approximation of density matrices}
\label{sec:approx_DM}

The case of a one-dimensional parameter space provides a simple intuitive way to illustrate a first version of the approximation method using Lagrange interpolation. We proceed therefore in two steps, explaining first the one-dimensional case before extending the methodology to higher-dimensional parameter spaces.

\subsubsection{One-dimensional parameter space}
\label{sec:one-dimens-param}
We predefine the offline-stage here in the sense that we choose $Q+1$  interpolation points $\sp_i$, $i=0,\ldots,Q$, and compute the corresponding density matrices $\sD_i=\sD_{\sp_i}$ at those points. We choose $\sp_0$ and consider the tangent space $\mathcal T_{\sD_0} \Gr$ as above. 
For the remaining $Q$ points $\sp_i\in\mathbb P$, we build the Lagrange basis functions $L_i:\mathbb P\subset \R\to\R$:
\begin{equation}
	L_i(\sp) = \frac{\prod_{j\neq i} (\sp-\sp_j)}{\prod_{j\neq i} (\sp_i-\sp_j)}.
\end{equation}

In the online-stage, for any new $\sp\in\mathbb P$ we build the following approximation, using $\Gamma_i=\GrLog(\sD_i)$,
\begin{equation}
	\Gamma(\sp) 
	= 
	\sum_{i=1}^Q L_i(\sp) \, \Gamma_i,
\end{equation}
upon which we apply the Grassmann exponential to finally obtain the approximate density matrix
\begin{equation}
	\sD_{\rm app}(\sp) = \GrExp\left( \sum_{i=1}^Q L_i(\sp) \, \Gamma_i \right).
\end{equation}
By construction, the interpolation property $\sD_{\rm app}(\sp_i) = \sD_i$ is satisfied due to the property $L_i(\sp_j)=\delta_{ij}$ of the Lagrange polynomials.

We note that when only two density matrices $\sD_0$ and $\sD_1$ are available, the application
\begin{equation}
	\sD_{\rm app}(\sp) = \GrExp\left( \frac{\sp-\sp_0}{\sp_1-\sp_0}  \Gamma_1 \right)
\end{equation}
parametrizes the geodesic between $\sD_0$ and $\sD_1$ on $\Gr$, as long as the exponential map is bijective, which is at least satisfied when $\sp_0$ and $\sp_1$ are close. 
This is the most natural way to define an approximation on $\Gr$ for values $\sp\in[\sp_0,\sp_1]$.

\subsubsection{Multi-dimensional parameter space}
\label{sec:multiD}

We now extend our considerations to arbitrary dimensional parameter domains. 
The previous case of a one-dimensional parameter space suggests that accurate approximations of $\Gamma$ can be obtained in the form of linear combinations of polynomials in $\sp$ times known vectors $\Gamma_i$ belonging to the tangent space.

We state now two remarks that seem appropriate at this point.

First, a possible generalization of the approach to higher dimensions can be realized by tensor-products of the Lagrange-polynomials. This would, however, require an exponential increase (with respect to the dimension) of data-points $\sp_i$ on a structured grid where the solutions $\sD_i$ and $\Gamma_i$, respectively, are required to be known. 
A remedy can consist of the use of sparse grids on the parameter domain but we will propose in the following a more adaptive framework.

Second, the set of all $\Gamma_i=\GrLog(\sD_i)$, $i=1,\ldots,Q$, might be highly linearly dependent. In such cases, there exists a low-dimensional basis $\{\Theta_1,\ldots,\Theta_n\}$, with $n \ll Q$, such that the manifold 
\begin{equation}
   \label{eq:DT0}
   \mathcal D_{0,\mathcal T} := \{ \Gamma(\sp) = \GrLog(\sDp) \,|\, \sp\in\mathbb P \}\subset \mathcal T_{\sD_0} \Gr,
\end{equation}
on $\mathcal T_{\sD_0} \Gr$ can be well-approximated by suitable elements of the $n$-dimensional space
\begin{equation}
   \label{eq:VRB}
   \mathbb V_{\rm rb} = \Span\{\Theta_1,\ldots,\Theta_n\}\subset \mathcal T_{\sD_0} \Gr,
\end{equation}
see Figure~\ref{fig:Gr} (right) for a schematic illustration of the situation. 
The approximate density matrix $\sD_{\rm app}(\sp)$  will be defined as $\sD_{\rm app}(\sp) := \GrExp\left( \Gamma_{\rm app}(\sp) \right)$, with
\begin{equation}
    \label{eq:gammaapp}
    \Gamma_{\rm app}(\sp) 
	= 
	\sum_{i=1}^n L_i(\sp) \, \Theta_i,
\end{equation}
where the functions $L_i:\mathbb P\to\R$ and the reduced basis $\{\Theta_1,\ldots,\Theta_n\}$ have to be appropriately chosen.
We focus for now on the practical aspects of the method. A more theoretical approach is presented in Appendix~\ref{sec:app}.

We start by choosing a rather large number $N_\sp$ of parameters $\sp\in\mathbb P$ (of the order 100 in our test cases), covering the parameter space $\mathbb P$ in a reasonable way. For example, one can take a uniform grid (as in our numerical tests) or (quasi-) random points on the parameter space~$\mathbb P$. 
Then, the offline part can be summarized by the following two main steps.

First, $\pd$ parameter points $\{\sq_1, \ldots, \sq_\pd\}$ among the $N_\sp$
points $\{\sp_1, \dots, \sp_{N_\sp}\}$ are selected, for which the density matrices are computed and, as well, their
Grassmann logarithms which we denote by $\{\Gamma(\sp_1), \ldots, \Gamma(\sp_\pd)\}$. Second, a reduced basis
$\{\Theta_1,\ldots,\Theta_n\}$ with $n\le d$ (hopefully $n\ll d$) of Grassmann
logarithms is computed using a singular value decomposition (SVD), from which
the functions $L_i(\sp)$ are also deduced.

More precisely, we first choose $d\in\mathbb N$ multivariate functions $\{P_1,
\ldots, P_d\}$ with $P_j: \mathbb P \rightarrow \mathbb R$ for $j$ from 1 to $d$.
For simplicity, we take all multivariate monomials on $\mathbb P$ of cumulative
degree up to $M$ with a total of $\pd$ monomials.
However, other choices for a basis are possible and 
do not change the substance of the method. 
We then assemble the matrix $\widetilde{P}\in \R^{N_\sp \times \pd}$ containing the values of these functions at the parameters $\{\sp_1, \ldots,
\sp_{N_\sp}\}$, i.e.
$\widetilde{P}_{i,j} = P_j(\sp_i)$. 

The main idea is to minimize the error between the exact and approximate Grassmann logarithms on these $N_\sp$ samples, i.e. solve
\begin{equation}
   \min_{\Theta \in \R^{d \times (\Nb\cdot N)}} \| \Gamma_{\rm train} - \widetilde{P} \Theta \|,
\end{equation}
where $\Gamma_{\rm train} \in \R^{N_\sp \times (\Nb\cdot N)}$ contains as rows the $\Gamma(\sp_i)$ reshaped in vectors, and where $\|\cdot \|$ is a suitable norm.
An approximate solution to this problem is found by 
selecting a square submatrix of $\widetilde{P}$ using the so-called \textit{maxvol} method as introduced in~\cite{gostz-maxvol-2008}. It finds a quasi-{dominant} square $\pd \times \pd$ submatrix denoted by $\widehat{P} \in \R^{\pd \times \pd}$ of $\widetilde{P}$ by selecting $d$ samples $\{\sq_i\}_{i=1}^\pd$. 
The approximate $\Theta$ is then written in the form
\begin{equation}
   \Theta = \widehat{P}^{-1} \widehat{\Gamma},
\end{equation}
where $\widehat{\Gamma}\in \R^{d \times (\Nb \cdot N)}$ contains as rows the
reshaped Grassmann logarithms $\Gamma(\sq_i)$. A great feature of this method
is that it requires only the computation of the density matrices for the
selected parameters $\{\sq_1, \ldots, \sq_\pd\}$ and not for all $N_\sp$ parameters.
At this stage, the Grassmann logarithm for a new parameter $\sp$ can be computed via
\begin{equation}
   \label{eq:first_approx_gamma}
   \Gamma_{\rm app}(\sp) 
	= 
	\sum_{i=1}^d \left[ P(\sp) \, \widehat{P}^{-1} \right]_i \Gamma(\sq_i),
\end{equation}
with $P(\sp) = (P_1(\sp), P_2(\sp), \ldots, P_\pd(\sp))$.

\begin{algorithm}[t]
   \caption{Offline stage}
   \label{alg:offline}
   \DontPrintSemicolon
   \KwData{Domain $\mathbb{P}$ of parameter $\sp$; $\pd$ multivariate
   monomials $\{P_i\}_{i=1}^\pd$; relative truncation threshold
   $\varepsilon$ for the SVD.}
   \KwResult{A reduced basis $(\Theta_1, \ldots, \Theta_n)$ along with its
   size $n$ and a $\pd \times n$ matrix $Z$, that define approximation
    in~\eqref{eq:gamma_approx}.
    \\
    \textbf{Total complexity:} $O((N_\sp+\Nb N) \pd^2 + \Nb^\beta \pd)$ if $N_\sp > \pd > n$ and $\Nb N \gg \pd$.
    }
   Define a uniform grid of $N_\sp$ points $\sp_j \in \mathbb{P}$ such that $N_\sp \ge \pd$.
   \emph{This requires $O(N_\sp)$ operations.}\;
   Compute the matrix $\widetilde{P}\in \R^{N_\sp \times \pd}$ given by
   $\widetilde{P}_{i, j} = P_j(\sp_i)$.
    \emph{Since all $P_j(\sp)$ are monomials, complexity is $O(N_\sp \pd)$.}
   \;
   Apply the \textit{maxvol} method to the matrix $\widetilde{P}$ to obtain $\pd$
   indices of rows $\{\piv(i)\}_{i=1}^n$ and compute the corresponding
   submatrix $\widehat{P}$.
   \emph{The number of operations is $O(N_\sp \pd^2)$.}\;
   Define the set $\{\sq_i\}_{i=1}^\pd$ such that $\sq_i = \sp_{\piv(i)}$.
   \emph{The operation count is $O(\pd)$.}\;
   For each $\sq_i$ define $\widehat{\Gamma}_{i, :}$ by reshaping the computed value
   of $\Gamma(\sq_i)$ into a $\Nb \cdot N$ row vector.
   \emph{The complexity is $O(\Nb^\beta \pd)$, where $\beta$ depends
   on the eigenvalue solver.}\;
   Compute the SVD of the matrix $\widehat{\Gamma}\in \R^{\pd \times (\Nb \cdot N)}$,
   truncate it to the rank-$n$ approximation $U_n S_n V_n$ such that
   $\sigma_{n+1}(\widehat{\Gamma}) < \varepsilon
   \sigma_1(\widehat{\Gamma})$.
   \emph{This step is done in $O(\Nb N \pd^2)$ operations.}\;
   Reshape each row of $n \times (\Nb \cdot N)$ factor $V_n$ into a
   corresponding $\Nb \times N$-matrix $\Theta_i$.
   \emph{No need to perform any operations, since reshape does not require any
   actions.}
   \;
   Output $\pd \times n$ matrix as the product $\widehat{P}^{-1}U_n S_n$.
   \emph{Inverting, multiplying and diagonal scaling in $O(\pd^3 + n \pd^2 + n
   \pd)$ operations.}
   \;
   Output the reduced basis $\{\Theta_1, \ldots, \Theta_n\}$.
\end{algorithm}

The second part consists of further reducing the dimensionality by performing a SVD on the matrix~$\widehat{\Gamma}$, noting that its rows can be highly linearly dependent. The SVD writes 
\begin{equation}
   \widehat{\Gamma} = \widehat{U}_n \widehat{S}_n \widehat{V}_n +
   \widehat{E}_n, \qquad
   \widehat{U}_n \in \R^{\pd \times n},
   \quad \widehat{S}_n \in \R^{n
   \times n},
   \quad \widehat{V}_n \in \R^{n \times (\Nb \cdot N)},
   \label{eq:svd}
\end{equation}
where $\widehat{E}_n$ is the remaining error term due to truncation.
The truncation order $n$ is determined based on a user-specified error tolerance $\varepsilon$ by requiring $\sigma_{n+1}(\widehat{\Gamma}) < \varepsilon \sigma_1(\widehat{\Gamma})$, where $\sigma_i(\widehat{\Gamma})$ denotes the $i$-th singular value of $\widehat\Gamma$.
We denote by $(\Theta_1, \ldots, \Theta_n)$ the rows of the matrix $\widehat{V}_n$ reshaped into matrices of size $\Nb \times N$.
Substituting the truncated SVD into \eqref{eq:first_approx_gamma} leads to
\begin{equation}
   \label{eq:gamma_approx}
   \Gamma_{\rm app}(\sp) 
	= 
	\sum_{i=1}^n \left[ P(\sp) \, Z \right]_i \Theta_i,
\end{equation}
where $Z = \widehat{P}^{-1} U_n S_n \in \R^{\pd \times n}$ and $\Theta_i$ can be precomputed offline. 
Thus, the online stage consists of building, for any new parameter $\sp\in\mathbb P$, the matrix $P(\sp)$, building $\Gamma_{\rm app}(\sp)$ according to~\eqref{eq:gamma_approx} and finally computing the Grassmann exponential thereof in order to obtain the approximate density matrix $\sD_{\rm app}(\sp) $.

The algorithms presenting the computations done in the offline and online
stages are described in Algorithm~\ref{alg:offline} and~\ref{alg:online},
together with the complexity of their different operations. Note that the most
time-consuming step in the online calculation is the application of the Grassmann exponential.

\begin{algorithm}[t]
   \caption{Online stage}
   \label{alg:online}
   \DontPrintSemicolon
   \KwData{A point $\sp\in\mathbb{P}$; $\pd$ multivariate monomials $\{P_i\}_{i=1}^\pd$;
   the reduced basis $\{\Theta_1, \ldots, \Theta_n\}$; 
   the matrix $Z \in\R^{\pd \times n}$ appearing in equation~\eqref{eq:gamma_approx}}
   \KwResult{The approximate value of $\sD(\sp)$  
   \\
   \textbf{Total complexity:} $O(\Nb N (n+N))$ if $\Nb N \gg \pd$.
}
    Compute the vector $P(\sp) = (P_1(\sp), \ldots, P_\pd(\sp))$ at the new point $\sp$.
   \emph{The number of operations is $O(\pd)$.}\;
    Compute the vector of scalars $(L_1(\sp), \ldots, L_n(\sp)) = P(\sp) Z$.
   \emph{This multiplication is done with $O(n \pd)$ operations.}\;
   Compute the matrix $\Gamma_{\sf app}(\sp) = \sum_{i=1}^n L_i(\sp) \Theta_i$.
   \emph{Summation with $O(n \Nb N)$ operations.}\;
   Apply Grassmann exponential: $\sD_{\sf app}(\sp) = \GrExp(\Gamma_{\sf
   app}(\sp))$
   \emph{Complexity of this step is mostly defined by the SVD leading to
   $O(\Nb N^2)$ operations}
 \end{algorithm}

\subsection{Summary of the method}
To summarize, the proposed approach returns an approximate density matrix $\sD_{\sf app}(\sp)$ of $\sD(\sp)$ 
at any given point $\sp$ inside the parameter domain $\mathbb{P}$. This density matrix $\sD_{\sf app}(\sp)$ is then used as an initial guess
for the SCF solver. The goal is to reduce the number of required SCF iterations. 
The starting guess is found with the two following steps:
\begin{enumerate}
    \item Offline, precomputations: define 
    points in
    $\mathbb{P}$ where the exact density matrix and functionals thereof are computed.
    \item Online, runtime computations: use the precomputed density matrices and functionals to reconstruct 
    an approximate density matrix $\sD_{\sf app}(\sp)$ at any parameter point
    $\sp \in \mathbb{P}$.
\end{enumerate}
The above mentioned steps are different for one-dimensional and multi-dimensional cases.
In the case of a one-dimensional domain $\mathbb{P}$, the data points are
chosen in a greedy hierarchical manner, as described in \cite{maday2009general}.
Then, a Lagrange interpolation is built upon these points. In the multi-dimensional case, we use
Algorithm~\ref{alg:offline}, performed offline, to obtain both the 
 points and the data.
Then, for any given value of $\sp\in\mathbb{P}$ we use Algorithm~\ref{alg:online} to compute an
initial guess. 

\section{Numerical results}
\label{sec:numerics}

To demonstrate the method's accuracy and robustness, we illustrate it on four different small- to medium-sized molecules, namely, the amino acids alanine, asparagine, phenylalanine, and tryptophan (13, 17, 23, and 27 atoms, respectively). 
If not explicitly stated otherwise, all the SCF calculations in the following have been performed using the CFOUR~\cite{cfour} suite of program, employing Dunning's cc-pVDZ basis set~\cite{Dunning_JCP_CCBasis}. The SCF program was modified so that a guess density matrix, obtained with the newly developed method, could be provided as an input. The default convergence criterion was used for all the calculation: $10^{-7}$ for the root-mean-square (RMS) change of the density and $10^{-6}$ for the maximum change.
The algorithm developed to generate the guess density, presented in Section~\ref{sec:approx_DM} has been implemented in Julia~\cite{bezanson_julia:_2017}. The program works with input densities which are generated by CFOUR, and writes as output the computed guess density matrix in a file, that can be read by CFOUR. 

In order to generate displaced geometries, normal modes are computed for the molecules using analytical second derivatives. For each molecule, we choose two different normal modes, one corresponding to the carbonyl C-O stretching, the second to a low-frequency collective vibration.
All the starting structures, including the normal modes used to generate displacement geometries, are provided in the supplementary material as input files for both CFOUR and PySCF.

As parameter values $\sp$, we consider the coefficients corresponding to each normal mode, i.e. the nuclear coordinates are constructed by
\begin{equation}
	\bm r = \bm r_0 + \sum_{i=1}^P \sp_i \bm n_i,
\end{equation}
where $\bm r_0 $ denotes the equilibrium geometry, $\sp_i$ the i-th component of the parameter $\sp$ and $\bm n_i$ the i-th normal mode.
For one-dimensional parameter domains, we consider thus one normal mode (the one reported first in the supporting information) whereas for two-dimensional domains, we consider both normal modes. The parameters $\sp_i$ are chosen in the range $[-1,1]$ bohr and discretized using an 11 points grid, i.e., we displace the geometries of $-1, -0.8, \ldots$ times the normal coordinate. For alanine, we repeat the calculations taking the larger parameters domain $[-10,10]$ bohr, still using an 11 points grid. The latter example is denoted by ``Alanine*''.
The grids for two-dimensional domains are formed by a tensor product of the one-dimensional grids.
For any given parameter $\sp$, the corresponding molecular geometry can then be generated and used for a SCF calculation. 

In the following, we provide several numerical tests. 
We illustrate how we can provide accurate initial density matrices for one-dimensional and two-dimensional parameter spaces. To assess the quality of the guess, we report the number of SCF iterations required to achieve convergence and we compare it with the number of iterations required starting from a guess obtained by diagonalizing the core Hamiltonian, which is the default guess in CFOUR. For 1D grids, we use the method presented in Section~\ref{sec:one-dimens-param} whereas for examples involving a two-dimensional parameter space, we use the general algorithm reported in Section~\ref{sec:multiD}.
Before proceeding with the numerical tests, we report in Table \ref{tab:scfstd} the number of SCF iterations needed to converge the Hartree-Fock equations, using different guess procedures, at the equilibrium geometry of the various test systems. The calculations were performed with different softwares, namely, CFOUR, Gaussian 16~\cite{g16} and PySCF 1.7~\cite{PYSCF} and are therefore not directly comparable.  However, they provide a qualitative estimate of the number of SCF iterations one can expect for such calculations and thus a benchmark for our algorithm. As convergence criteria are different in the various codes, we consider the SCF converged when the maximum variation of the density matrix between two subsequent iterations is smaller than 10$^{-6}$, as this information is reported in all codes used for the various calculations. Furthermore, in order to make the comparison more direct, we used in all calculations only the standard DIIS extrapolation. In particular, for the calculations performed with Gaussian 16, we disabled the use of the energy-based DIIS extrapolation.
\begin{table} 
 \caption{Number of SCF iterations required to achieve convergence (max change in the density smaller than $10^{-6}$) using different initial guesses. As the computations were carried out using different packages, that offer different SCF implementations, this cannot be considered an accurate comparison between the various guesses, but only a qualitative estimate of the number of required iterations. Note that all the calculations have been performed using standard DIIS extrapolation, with the same extrapolation space of 20 vectors. Core: diagonalization of the core Hamitlonian (with CFOUR). Harris: diagonalization of the Harris  functional (Gaussian 16). H\"uckel: using the extended H\"uckel method (PySCF). MinAO: start from a SCF calculation using a minimal AO basis set, which is then projected onto the chosen basis (PySCF). SAD: superposition of atomic densities (PySCF)\label{tab:scfstd}.}
 \centering
 \begin{tabular}{lcccc}
  \toprule
             & Alanine & Asparagine & Phenylalanine & Tryptophan \\
  \midrule
  Core       & 21      & 21         & 23            & 26         \\
  Harris     & 13      & 14         & 14            & 15         \\
  H\"uckel   & 16      & 17         & 17            & 18         \\
  MinAO      & 15      & 17         & 17            & 17         \\
  SAD        & 16      & 17         & 17            & 17         \\
  \bottomrule
 \end{tabular}
\end{table}

\subsection{One-dimensional parameter domains}
For this first batch of tests with $P=1$, we compute an approximation of the density matrix using the method presented in Section~\ref{sec:one-dimens-param} for every point in the parameter space (i.e., for each displaced geometry) and use it as a starting guess for a SCF calculation in CFOUR. We repeat such computations varying the order of interpolation, i.e., the number of precomputed densities used to build the guess. In order to select the interpolation points, we select them with a hierarchical greedy algorithm that chooses as next point the parameter value where the current approximation is worst, sometimes also referred to the \textit{magic points} (see~\cite{maday2009general}).
In this simple one-dimensional case, we consider the left-most, thus the smallest, parameter value as the root to build the tangent space. We observe numerically that all the results are independent on the choice of the root to build the tangent space, which is not obvious from the formulae.

\begin{figure} \centering
  \includegraphics[width=0.48\linewidth]{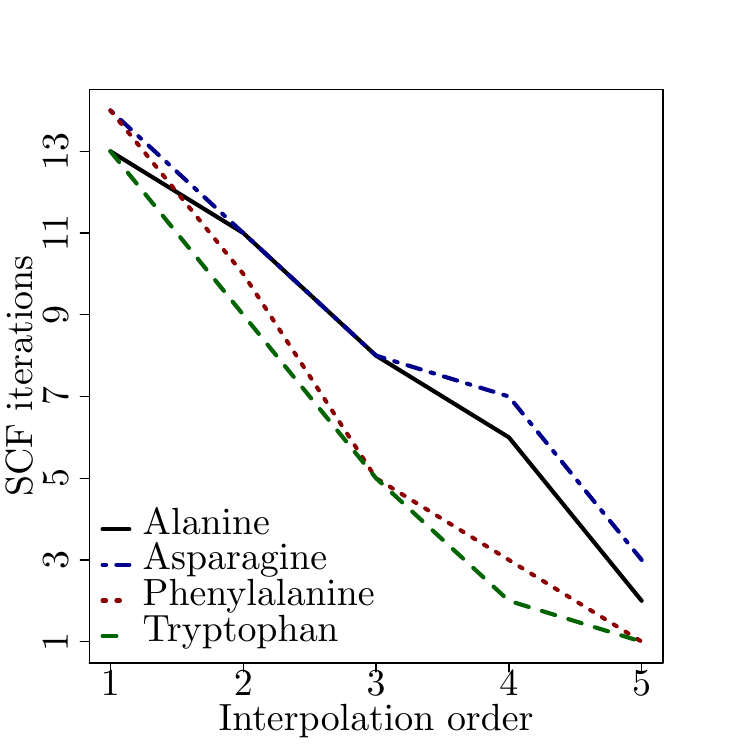}
  \includegraphics[width=0.48\linewidth]{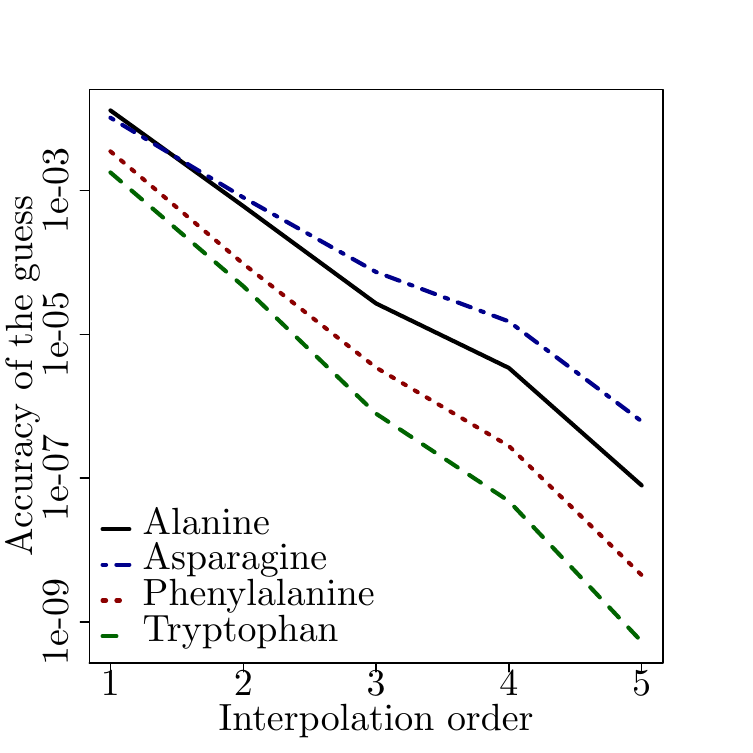}
  \caption{
    Results for the 1D parameter space. Number of SCF iterations required to achieve convergence (left panel) and Frobenius norm error on the density guess (right panel) as a function of the interpolation order for the various test systems. All the calculations were performed with CFOUR using the following convergence criteria for the increment of the density $\Delta P$: RMS $\Delta P < 10^{-7}$ and max $|\Delta P| < 10^{-6}$.}
  \label{fig:scf}
\end{figure}

The results obtained using our guessing procedure for the four amino acids selected as test molecules are reported in Figure~\ref{fig:scf}. In the left panel, we show the maximal number of SCF iterations required to achieve convergence over all the points in the test grids. In the right panel, the accuracy of the guess with respect to the converged SCF density is also reported. The tests confirm the good accuracy of our guess, as using a Lagrange polynomial interpolation of degree $5$ manages to reduce the number of required SCF iterations to only a few, namely, 3 for asparagine, 2 for alanine, and to 1 for phenylalanine and tryptophan. The latter result is particularly noteworthy as it demonstrates that, for these two systems, our guessing procedure can produce a guess density which is essentially already at convergence, as it can also be seen by looking at the norm of the error in the right panel. This makes in turn the overall SCF procedure unnecessary. For the other two molecules, convergence is achieved in 2 or 3 iterations, which is still a remarkable gain with respect to the standard procedure, that always requires at least 13 iterations. 

We point out that while we are considering small perturbations to the equilibrium geometry, these are not negligible. The SCF energy along the grid points varies of about 1.5-2.0 kcal/mol, which is a small, but significant oscillation if compared with the thermal energy at room temperature. In the following examples, we will explore larger energy fluctuations in order to assess the robustness of the method.

An interesting comparison can be made here with a common-practice strategy to provide a good guess for SCF calculations at similar geometries, i.e., using the converged SCF density as a guess for a calculation at a close geometry. We proceed as follows. We compute a fully converged SCF solution at the first gridpoint and then we advance along the 1D grid using each time the SCF density of the previous point as a guess.
Considering the 10 points for which a guess density was available, the SCF converged on average in 10 iterations for tryptophan, 11 iterations for alanine and asparagine, and 12 iterations for phenylalanine. While these numbers are, as it could be easily expected, an improvement with respect to the ones reported in Table \ref{tab:scfstd}, it is apparent how our algorithm outperforms this strategy. We also repeated the calculation for alanine on the coarser grid, i.e., using displacements of 1 bohr along the normal coordinate. In this case, 13-14 iterations were needed to achieve convergence, which is close to what reported in Table \ref{tab:scfstd}, meaning that the geometry change considered for this example is already more than enough to produce sizeable changes in the density matrix and hindering thus the efficiency of a simple strategy such as using the density at the closest available geometry.

\subsection{Two-dimensional parameter domains}

We now present similar tests for the case where the parameter domain is two-dimensional, i.e., we allow the displacement of the atoms in the molecules in two normal directions ($P=2$).
The initial guess density matrix is computed with the method presented in Section~\ref{sec:multiD}, using a maximum cumulative degree of the monomials taken to $M=8$ with a corresponding number of monomials $\pd=45$.
This ensures in the following numerical tests that the tolerances obtained in equation~\eqref{eq:svd} are reasonably small. 

\begin{figure}
    \centering
    \subfloat[$M=5$]{\includegraphics[width=0.35\textwidth]{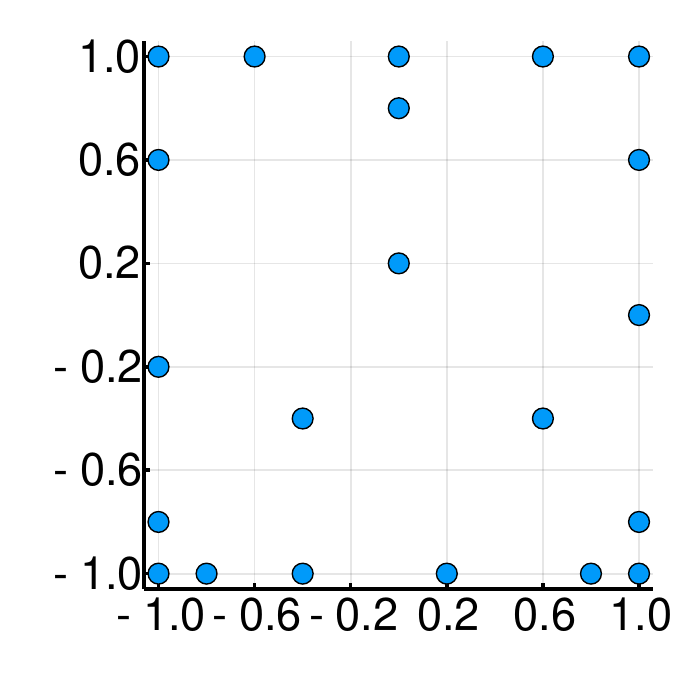}}
    \subfloat[$M=8$]{\includegraphics[width=0.35\textwidth]{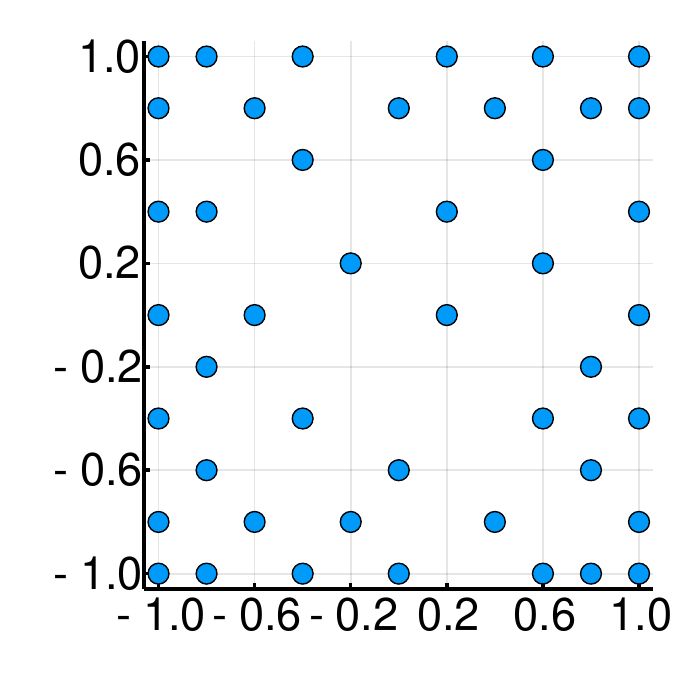}}
    \caption{Maxvol-selected points for 2D case for different maximum cumulative degree
    $M$.}
    \label{fig:maxvolpoints}
\end{figure}

For the two-dimensional grid used here, we generate a uniform $11 \times 11$ test-grid consisting of 121 points, i.e., displaced geometries. Note that in the offline part, the required SCF computations are only those of the selected parameters in the maxvol method presented in Section~\ref{sec:multiD}, i.e. only $\pd=45$ calculations. 
Figure~\ref{fig:maxvolpoints} shows the actual points selected by the
maxvol-algorithm, for maximum cumulative degrees $M$ equal to $5$ and $8$ respectively.
The converged density matrices at the selected points are then used to
build the reduced basis, the size of which is reported in the following. 
For the four considered molecules, using the $[-1,1]$ parameter range, the SCF energy exhibits much larger fluctuations than the ones observed for the $P=1$ examples. In particular, the energy fluctuates of 9.1, 8.9, 8.5, and 7.6 kcal/mol for alanine, asparagine, phenylalanine, and tryptophan, respectively. These are large energy fluctuations for a single molecule if compared, for instance, with the thermal energy at room temperature, and are likely not to be encountered when performing a molecular dynamics simulation. 
As in the 1D case, we chose the lower-left parameter value as the root to build the tangent
  space, and we observe numerically that the results are independent of the choice of the root to build the
  tangent space.

\begin{figure}
  \centering
  \includegraphics[width=0.48\linewidth]{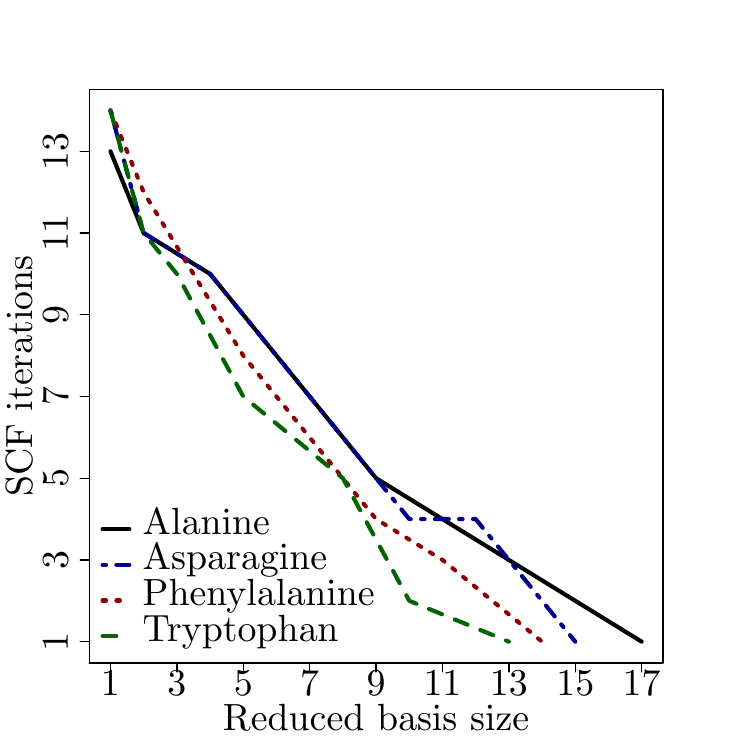}
  \includegraphics[width=0.48\linewidth]{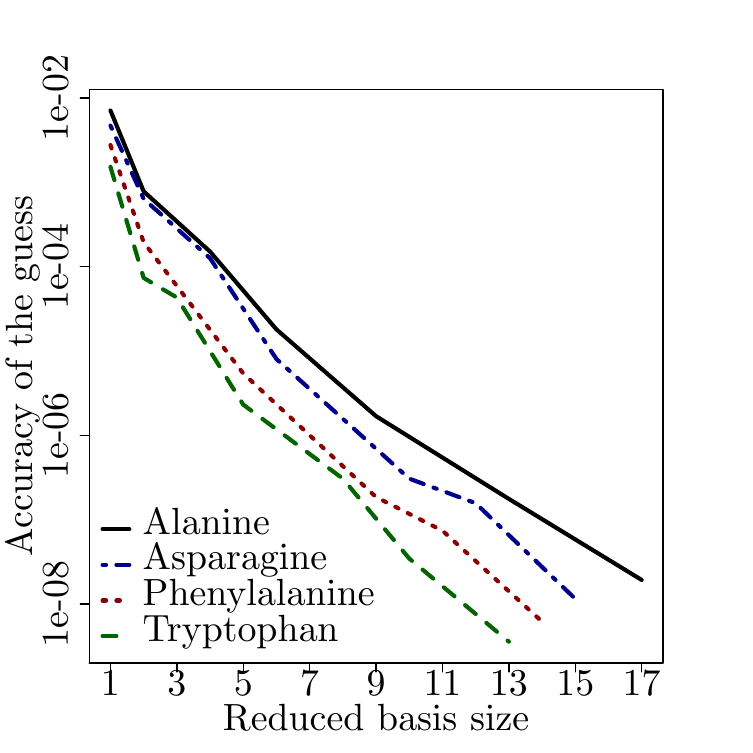}
  \caption{
   Results for the 2D parameter space. Number of SCF iterations required to achieve convergence (left panel) and Frobenius norm error on the density guess (right panel) as a function of the interpolation order for the various test systems. All the calculations were performed with CFOUR using the following convergence criteria for the increment of the density $\Delta P$: RMS $\Delta P < 10^{-7}$ and max $|\Delta P| < 10^{-6}$.}
  \label{fig:scf2d}
 \end{figure}

In Figure~\ref{fig:scf2d} (left panel) we report the maximum number of SCF iterations required to achieve convergence over the test grid of parameter values as a function of the size of the reduced basis used to build the approximation. In the right panel, the error of the computed guess with respect to the converged SCF density is reported. These results show that, despite the sizeable fluctuations in the energy, our procedure is always able to reconstruct a guess density that is at convergence for every displaced geometry using no more than 17 basis vectors, with 13 being enough to obtain the same result in the best-case scenario (tryptophan). The convergence of the error in the density with respect to the number of basis vectors (right panel) is fast and smooth, which confirms the excellent performances of our procedure.

A computational remark is, at this point, mandatory. The guess procedure presented in Section~\ref{sec:met} consists of two separate parts, named offline and online stages, respectively. In the offline part, the reduced basis is assembled. This is of course the expensive part of the procedure, as in order to compute the reduced basis, we need to solve the SCF problem at a given number of points, depending on the required accuracy. 
The online state is, on the other hand, completely inexpensive and can be performed in a fraction of a second for all the examples reported in this work. 
The key idea beyond the separation of the procedure in two different stages is that the offline one can be performed once and for all: as soon as the reduced basis is available, only the online stage has to be performed. In practice, this means that if we were to repeat our test calculations with a much finer grid, we would get a guess density for all the points using the reduced basis already assembled, and therefore at a cost that is completely negligible with respect to that of performing even a single SCF iteration.

\begin{figure}
   \centering
   \includegraphics[width=0.48\linewidth]{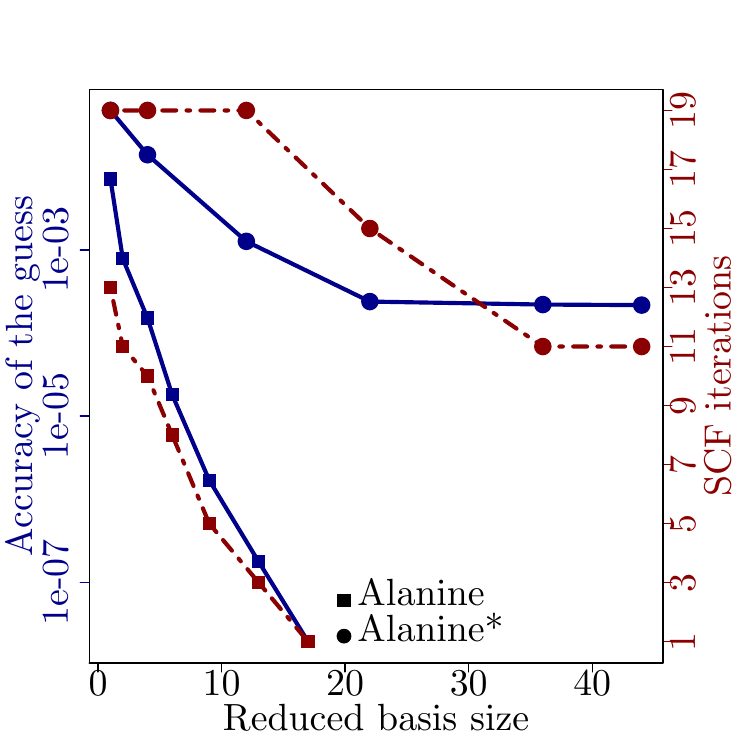}
   \includegraphics[width=0.48\linewidth]{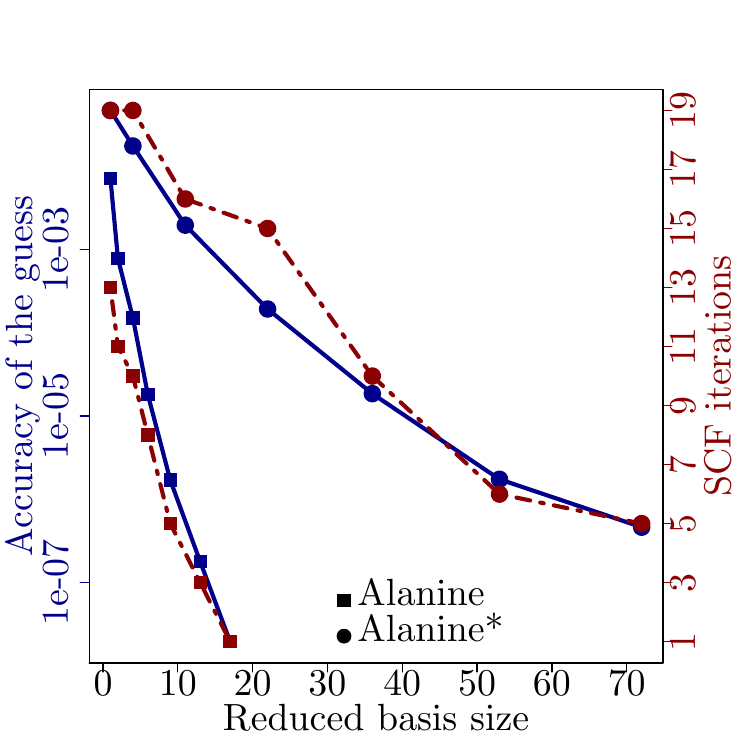}
   \caption{
    Comparison of alanine and alanine* in the 2D parameter space. Number of SCF iterations required to achieve convergence and Frobenius norm error on the density guess as a function of the interpolation order for $M=8$ for both systems (left panel) and $M=8$ for alanine and M=14 for alanine* (right panel). All the calculations were performed with CFOUR using the following convergence criteria for the increment of the density $\Delta P$: RMS $\Delta P < 10^{-7}$ and max $|\Delta P| < 10^{-6}$.}
   \label{fig:alanine}
 \end{figure}

In order to test the robustness of our procedure, we repeated the calculations on alanine using a two-dimensional grid, this time with a parameter domain of $[-10,10]$. This grid encompasses large geometry variations, with the SCF energy varying in a range of more than 1000 kcal/mol, and provides a test for our algorithm in more extreme conditions. In Figure~\ref{fig:alanine} we report, in the left panel, the results obtained for this case using the same setup used for the other 2D examples. 
The number of SCF iterations is reported on the right axis, while the error is on the left. For comparison, the results for the same molecule and the previous grid are always reported. 
We can immediately see how our guess procedure is now struggling to provide an accurate guess. Increasing the size of the reduced basis, we observe that the accuracy is stagnating, so that there is no gain by further increasing it. In order to better understand the source of this behavior, we allow the maximum cumulative degree of the monomials used in the algorithm to grow up to 14. The results are reported in the right panel of Figure~\ref{fig:alanine}. 
The guess density error and the number of SCF iterations exhibit now a convergent behavior, with as little as 5 iterations needed to converge the SCF in the worse case scenario when using the largest reduced basis. However, the size of the reduced basis required to observe a large reduction of the number of SCF iterations is much larger than what was observed before. 
We stress however that this is an extreme test case, and that we compute the SCF at geometries that are always quite distant from each other and it is hard to imagine a similar situation in a real-life application. However, the reduced basis built for this example allows one to explore a much larger portion of the potential energy surface of alanine than before, so that a larger number of vectors in the reduced basis appears justified. We stress that, even though the reduced basis is much larger than in the other examples, the online stage of the algorithm can still be performed in a negligible amount of time (less than 1 second). 

\begin{figure}
   \centering
   \includegraphics[width=0.48\linewidth]{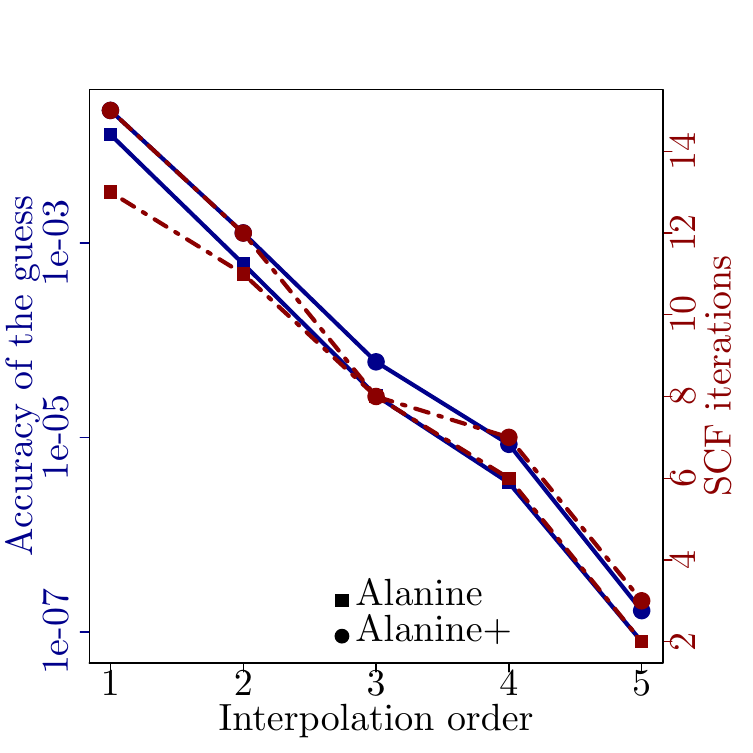}
   \includegraphics[width=0.48\linewidth]{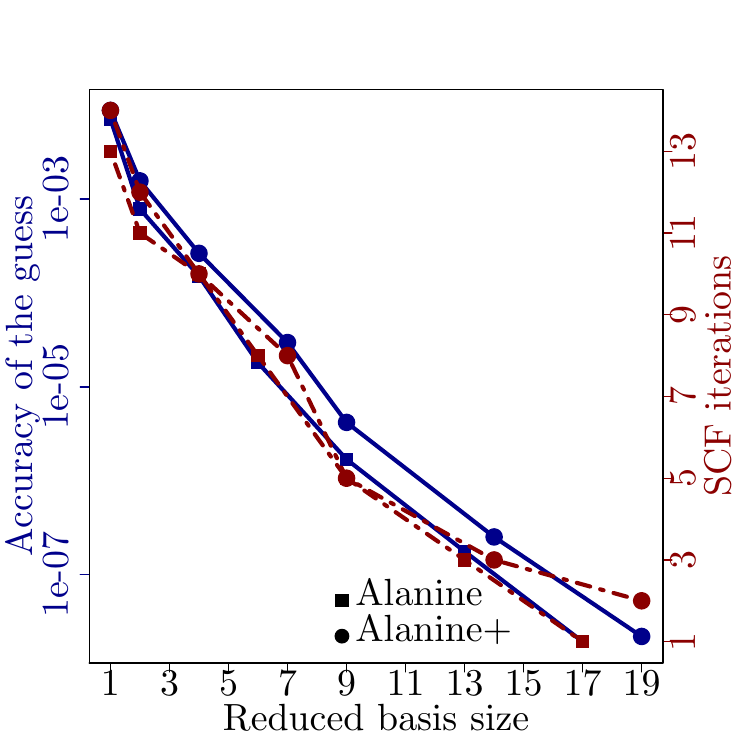}
   \caption{
  Comparison of alanine and alanine+. Number of SCF iterations required to achieve convergence and Frobenius norm error on the density guess as a function of the interpolation order for the 1D parameter space (left panel) and the 2D parameter space (right panel). All the calculations were performed with CFOUR using the following convergence criteria for the increment of the density $\Delta P$: RMS $\Delta P < 10^{-7}$ and max $|\Delta P| < 10^{-6}$.}
   \label{fig:alanine_aug}
\end{figure}

Finally, in order to check the method when a larger basis set is used, we repeated the calculations, once again chosing alanine and employing the fine 2-dimensional grid, using the augmented, triple zeta Dunning's basis set aug-cc-pVTZ. These sets of results are labeled ``Alanine+'' and reported in Figure \ref{fig:alanine_aug}, where they are compared vis-a-vis with the results obtained with the smaller cc-pVDZ basis set. As it can be seen from the figure, the use of a larger basis set has virtually no influence on our algorithm. This result is not surprising, as the methodology applies in principle to the non-discretized problem as well, i.e., for complete basis sets.

\section{Perspectives}
\label{sec:perspectives}
In this contribution, we presented a new method to compute a guess density matrix for the self-consistent field procedure at a given molecular geometry exploiting in an efficient way results available for other molecular geometries. The method is robust and is able to efficiently reconstruct a very good approximation to the SCF density at the a new geometry, often to the point that the SCF procedure itself becomes unnecessary. The proposed algorithm is divided in two different steps. 
In a first, offline phase, the building blocks for the approximation of the density are computed.
This phase thus encompasses all the most expensive steps in the calculation, including solving the SCF problem at a number of geometries, which are chosen using a greedy strategy that attempts to add, at every new point, the most relevant information to improve the basis. Once this first stage is completed, the online phase comes into play. Starting with the results of the offline stage, the approximated density is built for any new molecular geometry. The cost of this second phase is negligible and the computational investment of the offline phase can be harvested in a many-query context where the online phase is used many times or, in other words, the effort done to assemble the reduced basis pays off when many other computations need to be performed, as a very good guess can be assembled for all such computations at a very little cost. 

In this first work, we tested the algorithm on a few selected medium-sized molecules, namely, the amino acids alanine, asparagine, phenylalanine and tryptophan. In order to create displaced geometries, we computed normal modes and chose two particular vibrations, namely, the carbonyl stretching and a low-frequency collective mode, and used such coordinates \gd{to} create either one- or two-dimensional grids, displacing the equilibrium geometry of 5 uniform increments per direction per dimension, generating thus 11 and 121 geometries, respectively, for the 1D and 2D cases. 
We tested our method both with displacements compatible with steps used in finite difference calculations of energy gradients, for which we observed variations in the SCF energy of about 7-9 kcal/mol, and also for much larger displacements, that gave rise to a range of SCF energies spreading well over 1000 kcal/mol. In both cases and for all molecules, the algorithm showed very good performances, generating a guess able to reduce the number of SCF iterations required to achieve convergence to only a few, if any was needed at all.

The main limitation of our strategy is that, at the moment, it was tested and applied only to low dimensional problems (in parameter domain) - as these are the only ones for which it is possible to generate uniform grids and compute reference data at each point with reasonable computational resources. 
The next natural stage is to test the algorithm on a more general set of data for a high-dimensional parameter domain and develop a strategy to handle the creation of a reduced basis when there is no simple connection between the different geometries. That would be the case if the geometries were generated randomly or with molecular dynamics. The latter application is of course particularly interesting. However, further understanding of the theory is still required and a new strategy to assemble the reduced basis on-the-fly has to be developed in order to circumvent the so far artificial offline-online decomposition.

\section*{Acknowledgements}
This paper is dedicated to prof. J{\"u}rgen Gauss in honor of his sixtieth birthday. FL would like to express his deepest gratitude to prof. Gauss, not only for being an exceptional, dedicated mentor, who deeply cares for his pupils both from a scientific and a human point of view, but also for all the things done together, that range from going to the opera, to skiing, to enjoying a nice dinner and a good bottle of wine. Having the chance of working in J{\"u}rgen's group has made a big difference not only for my scientific growth, but also for my personal one. 

This work has been created based on an interdisciplinary collaboration between theoretical chemistry and applied mathematics. The mathematics community had the chance to meet and interact with prof. Gauss in interdisciplinary workshops, such as the Oberwolfach workshop on ``Mathematical Methods in Quantum Chemistry'' held in March 2018 for example, and learnt to know him as a researcher who is interested in fundamental concepts and answers to scientific questions, which of course is a common interest with mathematics. 
In this regards, we think that this article reflects this particular facet of prof. Gauss' research activities. 

The roots of this project lie in a project given to students in the so-called CAMMP Week Pro\footnote{https://blog.rwth-aachen.de/cammp/angebot-fuer-studierende/} (where CAMMP stands for Computational And Mathematical Modelling Program)
where students work during one week intensively in a team on a research-related problem. 
It was fascinating to see what students can achieve within one week and their work has definitively contributed to get the ball rolling in this project.

Finally, the authors acknowledge Eric Canc\`es and Yvon Maday for fruitful discussions.

\section*{Funding}

This work was supported by the French "Investissements d'Avenir" program, project ISITE-BFC (contract ANR-15-IDEX-0003).

\appendix
\section{Appendix}
\label{sec:app}

We describe in this appendix the method used in the case of multi-dimensional parameter spaces. In particular, we detail the motivations and justifications behind the choices leading to the method presented in Section~\ref{sec:multiD}.

First, the approximate density matrix $\sD_{\rm app}(\sp)$  will be defined as $\sD_{\rm app}(\sp) := \GrExp\left( \Gamma_{\rm app}(\sp) \right),$ with
\begin{equation}
    \label{eq:gammaapp1}
    \Gamma_{\rm app}(\sp) 
	= 
	\sum_{i=1}^n L_i(\sp) \, \Theta_i,
\end{equation}
where the functions $L_i:\mathbb P\to\R$ and the reduced basis $\{\Theta_1,\ldots,\Theta_n\}$ have to be appropriately chosen.

Since each $\Theta_i$ consists of $\Nb \cdot N$ elements, an obvious upper
bound on a dimensionality of the reduced basis is $n \le \Nb N$, but we hope to have a reduced basis of a much smaller i.e.
 $n \ll \Nb N$.
Let us assume  for now that some arbitrary set of functions
$\{L_i(\sp)\}_{i=1}^n$ are given and that they are stored in a row-vector
$L(\sp)=(L_1(\sp), L_2(\sp), \ldots, L_n(\sp))$.
Let us denote by $\Theta$ the 3-dimensional tensor such that $\Theta(i, :, :) = \Theta_i$ which is
simply a stack of all elements $\Theta_i\in \R^{\Nb\times N}$. Since we are looking for approximations of the
form \eqref{eq:gammaapp1}, we can rewrite it in compact notation as
\begin{equation}
    \Gamma_{\rm app}(\sp) = L(\sp) \Theta.
\end{equation}

With these considerations, it now becomes clear that we have to optimize
simultaneously the reduced basis $\mathbb V_{\rm rb}$ as well as the
functions $L_i(\sp)$ contained in the vector $L(\sp)$, i.e., we consider the minimization problem
\begin{equation}
    \label{eq:optim}
    \min_{\Theta\in\R^{n\times\Nb\times N}}
    \min_{L }  
    \Vert \Gamma(\cdot) - L(\cdot)  \Theta \Vert_*,
\end{equation}
where the norm $\Vert \cdot \Vert_*$ is arbitrary and any suitable norm can be
chosen.

This problem can also be viewed from a different angle: 
For the given $\Nb \cdot N$ functions $\Gamma_{j,i}(\sp)$ and given functions $L_1(\sp), L_2(\sp), \ldots,
L_n(\sp)$, one aims to approximate each $\Gamma_{j,i}(\sp)$ in the space
spanned by elements of~$L$, i.e., the $L_i$. 
Then, the ansatz~\eqref{eq:gammaapp1} can be seen as finding
coefficients $\Theta_1, \Theta_2, \ldots, \Theta_n$, thus the reduced basis,
for given row-vector $L(\sp)$.
This corresponds to exchanging the order of the minima in~\eqref{eq:optim}.

From this perspective we first prescribe $\pd$ polynomial basis functions
$P_1(\sp), P_2(\sp), \ldots,P_\pd(\sp)$, collected in the vector
$P(\sp)=(P_1(\sp), P_2(\sp), \ldots, P_\pd(\sp))$, spanning a sufficiently large space such that the distance between $\Gamma(\sp)$ and its projection to the space
spanned by $P(\sp)$ is smaller than a certain
threshold.
Just like the size $n$ of the reduced basis, reasonable value of $\pd$ is
assumed to satisfy $\pd \ll \Nb N$.
Then, for given functions $P_i(\sp)$ (and thus $P(\sp)$) one is aiming at a $\Theta$ that minimizes the following distance:
\begin{equation}
    \label{eq:argminfrobenius}
    \Theta_{P} 
    = 
    \argmin_{\Theta\in\R^{\pd\times\Nb\times N}} \Vert \Gamma(\cdot) -
    P(\cdot)  \Theta \Vert_*.
\end{equation}
Note that the dimension $\pd$ of the reduced basis, as constructed like this, will be reduced in a further step.
As norm $\|\cdot\|_*$, we will first consider the ideal choice
\begin{equation}
    \Vert \Gamma(\cdot) -  P(\cdot) \Theta \Vert_*^2 
    = \int_{\mathbb P} \Vert
    \Gamma(\sp) -  P(\sp) \Theta \Vert_F^2  \,d\sp,
\end{equation}
as starting point.
Here $\Vert \cdot \Vert_F$ stands for the Frobenius norm for matrices.
Having the exact $\Gamma(\sp)$ at every point $\sp \in \mathbb P$ is not feasible in practice which motivates to introduce 
a quadrature rule based on points~$\sp_j$ and weights $\omega_j$,
$j=1,\ldots,N_\sp$ given by:
\begin{equation}
	\label{eq:circnorm}
   \Vert \Gamma(\cdot) -  P(\cdot) \Theta \Vert_\circ^2 
    :=
    \sum_{j=1}^{N_\sp} w_j
     \Vert \Gamma(\sp_j) -  P(\sp_j) \Theta \Vert_F^2
     \approx 
      \Vert \Gamma(\cdot) -  P(\cdot) \Theta \Vert_*^2 .
\end{equation}
Introducing $\widetilde\Gamma\in\R^{N_\sp\times(\Nb\cdot N)}$,
$\widetilde\Theta\in\R^{\pd\times(\Nb\cdot N)}$ and $\widetilde{P}\in\R^{ N_\sp \times \pd}$
defined by
\begin{align}
	\widetilde\Gamma_{j,:} &= \mathsf{reshape}(\Gamma(\sp_j),1,\Nb\cdot N),\\
        \widetilde\Theta &= \mathsf{reshape}(\Theta,\pd,\Nb\cdot N),\\
	\widetilde P_{j,i} &= P_i(\sp_j),
\end{align}
we rewrite the optimization problem as follows:
\begin{equation}
    \label{eq:argmintheta}
    \widetilde{\Theta}_{\widetilde{P}} =  \argmin_{\widetilde{\Theta}\in\R^{\pd\times(\Nb\cdot
    N)}}
    \Vert \widetilde{\Gamma} -
    \widetilde{P} \widetilde{\Theta} \Vert_F,
\end{equation}
assuming $\sp_j$ is a uniform grid in $\mathbb{P}$, i.e. $\omega_j=\frac{\vert
\mathbb{P} \vert}{N_\sp}$. 

In consequence, we transformed the problem to a least squares
problem, whose solution, for given $\widetilde P$, is given by the pseudoinverse of $\widetilde{P}$ acting on~$\widetilde \Gamma$:
\begin{equation}
    \label{eq:optimaltheta}
    \widetilde{\Theta}_{\widetilde{P}} = \widetilde{P}^\dagger \widetilde{\Gamma}.
\end{equation}
Thus, in the case where the matrix $\widetilde{P}$ is given, we have an explicit expression for the minimizer and one can easily compute the optimal coefficients $\Theta_i$ of the approximation~\eqref{eq:gammaapp}. 
Returning to the global optimization problem~\eqref{eq:optim}, this allows us to write the optimization problem in only one variable, namely the $N_\sp \times \pd$ matrix $\widetilde{P}$. 
The minimization problem becomes
\begin{equation}
    \widetilde{P}_{\sf opt} = \argmin_{\widetilde{P} \in \R^{N_\sp \times \pd}}
    \Vert \widetilde{\Gamma} - \widetilde{P} \widetilde{P}^\dagger
    \widetilde{\Gamma}\Vert_F.
\end{equation}
The solution of such an optimization problem is the best approximation of $\widetilde{\Gamma}$ by
matrices of (given) rank $\pd$ (the size of reduced basis) and can be obtained by performing the singular value decomposition of the
matrix $\widetilde{\Gamma} = U\Sigma V^\top$, so that $\widetilde{P}_{\sf opt}$
and $\widetilde \Theta_{\sf opt}$ are given by
\begin{equation}
    \label{eq:pthetaopt}
    \widetilde{P}_{\sf opt} = U_\pd, \qquad 
    \widetilde \Theta_{\sf opt} =
    U_\pd^\top \widetilde{\Gamma},
\end{equation}
where $U_\pd\Sigma_\pd V_\pd^\top$ is the rank $\pd$ approximation of
$\widetilde{\Gamma} $ with $U_\pd\in\R^{N_\sp\times \pd}$, $\Sigma_\pd\in\R^{\pd\times
\pd}$ and $V_\pd\in \R^{\Nb\cdot N\times \pd}$.
This provides the solution to the optimization problem~\eqref{eq:optim} for the particular norm defined in~\eqref{eq:circnorm}.

Unfortunately, the optimal $\widetilde{\Theta}_{\sf opt}$, as can be found in
equation~\eqref{eq:pthetaopt}, requires
full knowledge of $\widetilde{\Gamma}$, i.e., $\Gamma(\sp_j)$ for every
quadrature point $\sp_j$. In the following we show how we can drastically
reduce the amount of points $\sp$ where we need to compute $\Gamma(\sp)$. 
In order to do so, we propose to replace the
Frobenius norm in favour for the \textbf{max} norm for matrices, also known as Chebyshev norm, given by
\begin{equation}
	\| A \|_{\rm C} = \max_{ij} |a_{ij}|.
\end{equation}
This leads to the following optimization problem
\begin{equation}
    \widetilde{\Theta}_{\sf opt,{\rm C}} = \argmin_{\widetilde{\Theta} \in \R^{\pd \times
    (\Nb \cdot N)}} \Vert \widetilde{\Gamma} - \widetilde{P} \widetilde{\Theta}
    \Vert_{\rm C}.
\end{equation}
We then aim to find quasi-optimal solutions of this problem by so-called \emph{interpolative}
approximations of the form
\begin{equation}
    \label{eq:crossapp}
    \widetilde{\Gamma}_{\sf app} = C U R,
\end{equation}
where either $C$ is a collection of ``basis'' columns of $\widetilde{\Gamma}$
or $R$ is a collection of ``basis'' rows of $\widetilde{\Gamma}$ and $U$ is a ``core''
matrix.
If both $C$ and $R$ are submatrices of $\widetilde{\Gamma}$, then
the ``core'' matrix is, usually, an inverse or pseudo-inverse of
the intersection of the ``basis'' rows $R$ and the ``basis'' columns~$C$.
Such an approximation is called cross approximation since the  intersection of columns and rows reminds of a cross. A
theoretical analysis of cross approximations, provided in \cite{gt-maxvol-2001,
gostz-maxvol-2008},
proves that such a rank $\pd$ decomposition exists, i.e., $U\in \mathbb R^{\pd\times\pd}$, such that
\begin{equation}
    \label{eq:curoptimal}
    \Vert \widetilde{\Gamma} - C U R \Vert_{\rm C} \le (\pd+1) \, 
    \sigma_{\pd+1}(\widetilde{\Gamma}).
\end{equation}
Here, $\sigma_{\pd+1}(\widetilde{\Gamma})$ denotes the $(\pd+1)$-st singular value of $\widetilde{\Gamma}$ in descending order.
More recent results on the error estimation in the Chebyshev norm can be found in
\cite{mo-rectmaxvol-2018, oz-projmaxvol-2018}.
Although the Chebyshev norm is studied well in terms of
theory and practical methods, building cross approximations with controlled
error in the spectral or Frobenius norm is still ongoing research.
We refer to the recent papers \cite{zo-skeletonfrobenius-2018, ck-curfrobenius-2019} for
further information.

Since we are looking for the interpolative approximation by rows of
$\widetilde{\Gamma}$, the matrix $C$ appearing in
equation~\eqref{eq:crossapp} can be chosen arbitrary as long as the space
spanned by its columns approximates columns of $\widetilde{\Gamma}$ with
high enough precision. 
The best choice is, of course, the first left singular vectors of
$\widetilde{\Gamma}$, which again requires the undesirable full knowledge of
$\widetilde{\Gamma}$.
However, any column of $\Gamma(\sp)$ can by construction be well approximated
by an element in the space spanned by the elements of $P(\sp)$ (the polynomial basis functions),
so the matrix $C$ can be defined as the matrix $\widetilde{P}$, previously
defined as the vector $P(\sp)$ at all quadrature points $\sp_j$.
Then the ``core'' matrix $U$ is simply an inverse
of some submatrix of $\widetilde{P}$ and the matrix $R$ is just a collection of~$\pd$
rows of the matrix $\widetilde{\Gamma}$, corresponding to~$\pd$ computations of
$\Gamma(\sp)$.

This is realized by analyzing only the matrix $\widetilde{P}$ to select a few samples $\{\sq_j\}_{j=1}^\pd$ where we need to compute subsequently the matrices $\Gamma(\sq_j)$.
For this purpose we use the so-called \textit{maxvol} method as introduced in~\cite{gostz-maxvol-2008}.
It finds a quasi-\emph{dominant} square $\pd \times \pd$ submatrix of $\widetilde{P}$ in $O(N_\sp \pd^2)$ operations.
A $\pd \times \pd$ submatrix of the $N_\sp \times \pd$ matrix $\widetilde{P}$
is called \emph{dominant} if the modulus of its determinant does not grow if we
change one of its rows by any other row of $\widetilde{P}$. \emph{Quasi-dominance}
means that the modulus of the determinant does not grow by more than a factor of $1+\alpha$
with a small value of $\alpha$. Such a property is necessary for the theoretical
error estimation presented in equation~\eqref{eq:curoptimal}.

In practise, the \textit{maxvol} method takes the matrix $\widetilde{P}\in\R^{N_\sp\times \pd} $ as input and returns
the square quasi-dominant submatrix $\widehat{P}\in\R^{\pd\times \pd}$ along with a
matrix of coefficients $C\in\R^{N_\sp\times \pd}$, such that the product of
the coefficients by the submatrix is equal to the input $\widetilde{P}$, i.e.,
\begin{equation}
    \widetilde{P} = C \widehat{P}, \qquad C = \widetilde{P} \widehat{P}^{-1}.
\end{equation}
Let us denote $\{\piv(i)\}_{i=1}^\pd$ the set of row-indices such that $\widehat{P}_{i,j}=\widetilde P_{\piv(i),j}$.
We now define
\begin{equation}
    \widehat{\Gamma}_{i,j}=\widetilde \Gamma_{\piv(i),j}
    \qquad
    \sq_i = \sp_{\piv(i)}.
\end{equation}
Then, the interpolative approximation of $\widetilde{\Gamma}$ is given by 
\begin{equation}
	\label{eq:curgamma}
        \widetilde \Gamma \approx \widetilde{P} \widehat{P}^{-1}
        \widehat{\Gamma} =
	\widetilde P \widetilde \Theta
\end{equation}
with $ \widetilde \Theta =\widehat{P}^{-1} \widehat{\Gamma}$ and we define the approximation
\begin{equation}
    \label{eq:maxvolinterpolation}
    \Gamma_{\rm app}(\sp) = \sum_{i=1}^\pd \left(\mathcal P(\sp) \widehat{P}^{-1}\right)_i
    \Gamma(\sq_i).
\end{equation}
One of the main features of this approximation is that, in order to compute the value of
$\Gamma(\sp)$ at a new point $\sp$, we only need to compute a row-vector
$P(\sp)$ of values of the polynomials $P_i(\sp)$ at the new point $\sp$ and that the functions 
$ \left(P(\sp) \widehat{P}^{-1}\right)_i$ are polynomials in the prescribed space.

Since we are working with $\widetilde P$ instead of $\widetilde\Gamma$,
the  actual error is different from the
estimation in \eqref{eq:curoptimal}. We omit the error analysis of our
approximation in this paper and plan to release it in a follow-up article.

Note that the ``basis'' rows
$\widehat{\Gamma}$ of the matrix $\widetilde{\Gamma}$ can be highly linearly
dependent. 
We therefore consider the singular value decomposition of the matrix
$\widehat{\Gamma}$ and truncate it up to rank $n$:
\begin{equation}
    \widehat{\Gamma} = \widehat{U}_n \widehat{S}_n \widehat{V}_n +
    \widehat{E}_n,
\end{equation}
such that $\widehat{U}_n \in \R^{\pd \times n}, \widehat{S}_n \in \R^{n
\times n}, \widehat{V}_n \in \R^{n \times (\Nb \cdot N)}$ and $\widehat{E}_n$
is the remaining term due to truncation. 
Substituting the truncated SVD into \eqref{eq:curgamma} we get:
\begin{equation}
    \label{eq:reshapedtruncatedapprox}
    \widetilde{\Gamma} \approx \widetilde{P} \widehat{P}^{-1} \widehat{U}_n
    \widehat{S}_n \widehat{V}_n.
\end{equation}
Let us denote the $i$-th row of $\widehat{V}_n$, after reshaping into a
$\Nb \times N$ matrix, as $\Theta_i$ and the product $\widehat{P}^{-1}
\widehat{U}_n S_n$ as a matrix $Z$.
Then, we approximate the value of $\Gamma(\sp)$ for any new value of $\sp$ as
\begin{equation}
    \label{eq:maxvolapproximation}
    \Gamma_{\rm app}(\sp) 
    = 
    \sum_{i=1}^n 
    L_i(\sp)
    \Theta_i.
\end{equation}
with $L_i(\sp) = \left( P(\sp) Z \right)_i$, which is exactly of the form~\eqref{eq:gammaapp1}.
The additional SVD further reduces the dimension of the reduced basis but of course introduces another error, which is nevertheless controlled by the singular values, but such an approximation requires a proper theoretical analysis, which is omitted in
this paper. It can be derived in the same way as the theoretical estimations in \cite[Theorem 4.8]{mo-rectmaxvol-2018}.

The proposed approximation technique can be formally divided into two
parts: an offline stage, where the reduced basis is pre-computed, and an online
stage, where an approximate value of $\sD(\sp)$ is computed efficiently for a given
$\sp$.

The offline part is schematically illustrated in Algorithm~\ref{alg:offline}. 
This stage requires $O((N_\sp+\Nb N)\pd^2 + \Nb^\beta \pd)$
operations, where $N_\sp$ stands for the number of quadrature points $\sp_j$, $\Nb$
and $N$ are the number of atomic orbital basis functions and the number of orbitals respectively.
Further, $\pd$ is the number of basis multivariate monomials and $\beta$ is a power factor determined by the specific nature of the eigenvalue solver that is employed to solve~\eqref{eq:evp1}-\eqref{eq:evp2}.

The online part containing the approximation of $\sD(\sp)$ for any new $\sp$ is sketched by Alg.~\ref{alg:online}.
It shall be noticed that this part is of much lower complexity: it uses only
$O(n(\Nb N + \pd^2) + \Nb N^2)$ operations, where $n$
is a size of the final reduced basis. 
It is worth to emphasize, that both procedures, offline as well as online, must use the same set of basis
monomials $\{P_1, \ldots, P_\pd\}$.

\bibliographystyle{tfo}

\end{document}